\documentclass[11pt]{article} % arXiv投稿時はdvipdfmxを削除
\usepackage{typearea}\typearea{12}
\usepackage{bm}
\usepackage{ascmac}
\usepackage{enumerate}
\usepackage[most]{tcolorbox}
\usepackage{amsmath, amssymb, amsfonts}
\usepackage{tikz, tikz-3dplot}
\usetikzlibrary{math}
\usepackage{caption}
\numberwithin{equation}{section}
\usepackage{mathtools, mathrsfs}
\usepackage{hyperref}
\hypersetup{hidelinks}
\usepackage{cite, url}

\def\eq#1\en{\begin{equation}#1\end{equation}}  
\def\eqa#1\ena{\begin{align}#1\end{align}}
\def\eqg#1\eng{\begin{gather}#1\end{gather}}
\newtcolorbox{summary}{sharp corners, left=-2.0mm, boxrule=0pt, colback=gray!35}

\newcommand{\Supp}{\operatorname{Supp}}
\newcommand{\Wid}{\operatorname{Wid}}
\renewcommand{\i}{\mathrm{i}}

\renewcommand{\a}{\alpha}
\newcommand{\ao}{\overline{\alpha}}
\renewcommand{\b}{\beta}
\newcommand{\bo}{\overline{\beta}}
\newcommand{\s}{\sigma}
\newcommand{\so}{\overline{\sigma}}
\renewcommand{\t}{\tau}
\renewcommand{\to}{\overline{\tau}}
\newcommand{\p}{\prime}
\newcommand{\calS}{\mathcal{S}}
\newcommand{\calP}{\mathcal{P}}
\newcommand{\calPL}{\mathcal{P}_{\Lambda}}
\renewcommand{\L}{\Lambda}
\newcommand{\ho}{\hat{1}}
\newcommand{\hA}{\hat{A}}
\newcommand{\hB}{\hat{B}}
\newcommand{\hC}{\hat{C}}
\newcommand{\hH}{\hat{H}}

\newcommand{\Hhop}{\hat{H}_{\mathrm{hop}}}
\newcommand{\Hint}{\hat{H}_{\mathrm{int}}}

\newcommand{\hQ}{\hat{Q}}

\renewcommand{\c}{\hat{c}}
\newcommand{\cp}{\hat{c}^{+}}
\newcommand{\cm}{\hat{c}^{-}}
\newcommand{\n}{\hat{n}}
\newcommand{\kb}{k}

\newcommand{\sfA}{\hat{\mathsf{A}}}
\newcommand{\sfB}{\hat{\mathsf{B}}}
\newcommand{\sfC}{\hat{\mathsf{C}}}
\newcommand{\sfD}{\hat{\mathsf{D}}}
\newcommand{\sfE}{\hat{\mathsf{E}}}
\newcommand{\sfF}{\hat{\mathsf{F}}}
\newcommand{\sfG}{\hat{\mathsf{G}}}

\newcommand{\qA}{q_{\sfA}}

\newcommand{\qSA}{q_{\calS(\sfA)}}

\newcommand{\x}{\mathrm{x}}
\newcommand{\y}{\mathrm{y}}
\newcommand{\xr}{\mathrel{\mathop{x}\limits^{\smash{\raisebox{-0.5ex}{$\scriptstyle\rightharpoonup$}}}}} % \right-most site
\newcommand{\xl}{\mathrel{\mathop{x}\limits^{\smash{\raisebox{-0.5ex}{$\scriptstyle\leftharpoonup$}}}}} % \left-most site
\newcommand{\up}{\uparrow}
\newcommand{\down}{\downarrow}

\newcommand{\ei}{\bm{e}_{1}}

\newtheorem{theorem}{Theorem}[section]
\newtheorem{lemma}[theorem]{Lemma}
   
\renewenvironment{theorem}
	{\medskip\refstepcounter{theorem}%
	\begin{itembox}[l]{\textbf{Theorem~\thetheorem}}
	\em}
	{\end{itembox}\medskip}

\renewenvironment{lemma}
	{\medskip\refstepcounter{theorem}%
	\begin{itembox}[l]{\textbf{Lemma~\thelemma}}
	\em}
	{\end{itembox}\bigskip}

\begin{document}

\tdplotsetmaincoords{65}{25} % 3D視点の角度設定

\noindent
{\Large \bf
Absence of nontrivial local conserved quantities in the Hubbard model on the two or higher dimensional hypercubic lattice
}

\renewcommand{\thefootnote}{\fnsymbol{footnote}}
\medskip\noindent
\medskip\noindent
Mahiro Futami\footnote{%
mahiro.futami@gmail.com
}\\
\medskip\noindent
{\footnotesize \sl
Department of Physics, Gakushuin University, Mejiro, Toshima-ku, Tokyo 171-8588, Japan
}
\renewcommand{\thefootnote}{\arabic{footnote}}
\setcounter{footnote}{0}

\begin{center}
\begin{minipage}{0.85\linewidth}
\small\noindent
By extending the strategy developed by Shiraishi in 2019, we prove that the standard Hubbard model on the $d$-dimensional hypercubic lattice with $d\ge2$ does not admit any nontrivial local conserved quantities.
The theorem strongly suggests that the model is non-integrable.
To our knowledge, this is the first extension of Shiraishi's proof of the absence of conserved quantities to a fermionic model.
Although our proof follows the original strategy of Shiraishi, it is essentially more subtle compared with the proof by Shiraishi and Tasaki of the corresponding theorem for $S=\tfrac12$ quantum spin systems in two or higher dimensions; our proof requires three steps, while that of Shiraishi and Tasaki requires only two steps.
It is also necessary to partially determine the conserved quantities of the one-dimensional Hubbard model to accomplish our proof.
\end{minipage}
\end{center}

%%%%%%%%%%%%%%%%
\tableofcontents
%%%%%%%%%%%%%%%%

%%%%%%%%%%%%%%%%
\section{Introduction}
\label{s:intro}

The Hubbard model, a tight-binding electron model with on-site interactions, is a standard idealized model of strongly interacting electrons in a solid.
It exhibits (or is expected to exhibit) rich phenomena, including antiferromagnetism, ferromagnetism, the Fermi liquid, and superconductivity.
See, e.g., \cite{Lieb1995,Tasaki1995,Arovas}.
In one dimension, the Hubbard model was solved by Lieb and Wu in 1968 using the Bethe ansatz method \cite{LiebWu1968,LiebWu2003}.
A series of exact conserved quantities was then discovered by Shastry \cite{Shastry1986, Shastry19867, Shastry1988}.
By now, there is almost complete understanding of conserved quantities in the one-dimensional Hubbard model \cite{Grosse1989, Yang1989, GrabowskiMathieu1994,Fukai2023,Fukai2024}.
Here, we shall show that the situation is essentially different in dimensions two or higher.

In 2019, Shiraishi developed a new method and proved that the spin-$\tfrac12$ XYZ chain under a magnetic field admits no nontrivial local conserved quantities \cite{Shiraishi2019}. 
Since integrable systems are typically characterized by the existence of infinitely many local conserved quantities, this result provides strong evidence that the model in question is not exactly solvable by conventional means. 
Shiraishi's method was subsequently extended to various one-dimensional quantum spin systems \cite{Chiba2023, ParkLee2025, Shiraishi2024, ParkLee2024,  HokkyoYamaguchiChiba2024, YamaguchiChibaShiraishi2024a,  YamaguchiChibaShiraishi2024b, Shiraishi2025}. 
Recently, strong results for the absence/presence of nontrivial local conserved quantities for general classes of one-dimensional quantum spin systems were developed in \cite{Hokkyo2025,ShiraishiYamaguchi2025}.

Shiraishi's method for proving the absence of nontrivial local conserved quantities was also extended to quantum spin models in two or higher dimensions by Shiraishi and Tasaki \cite{ShiraishiTasaki2024}, who worked on the XY and the XYZ models, and Chiba \cite{Chiba2024}, who worked on the quantum Ising models.
See also \cite{FutamiTasaki2025} for a similar result for the quantum compass model.

\medskip
In the present work, we extend Shiraishi's method to the Hubbard model in two or higher dimensions and establish that the model admits no nontrivial local conserved quantities.
This provides strong support for the common belief that the model is integrable only in one dimension.
We also stress that, to our knowledge, this is the first extension of Shiraishi's proof of the absence of conserved quantities to a fermionic model.
We believe that our method can be generalized to cover various fermionic models of physical importance.

One might suspect that the absence of nontrivial local conserved quantities of the Hubbard model in two or higher dimensions may be proved by a faithful modification of the corresponding proof in \cite{ShiraishiTasaki2024} for the $S=\tfrac12$ XYZ model (and that was what we expected in the beginning of the research).
It turns out, however, that this is not the case.
There is an essential difficulty intrinsic to the Hubbard model, which requires us to perform an extra analysis not present in \cite{ShiraishiTasaki2024}.
In short, our proof consists of ``three steps'' while that in \cite{ShiraishiTasaki2024} requires only ``two steps''.
See the end of Section~\ref{s:strategy} and also Section~\ref{s:discussion} for more details.

\section{Definitions and the main theorem}
\label{s:def}

Let $ \Lambda = \{ 1 , \ldots , L \} ^ { d } $ be the $ d $-dimensional $ L \times \cdots \times L $ hypercube lattice with periodic boundary conditions, where $ d \geq 2 $.
For a nonempty subset $ S \subset \L $, we define its width, denoted as $ \Wid S $, as the minimum $ w $ such that
\begin{align}
	0 \leq ( x ) _ { 1 } - a \leq w - 1 \ ( \text{mod } L ) ,
\end{align}
for every $ x \in S $ with some $ a \in \{ 1 , 2 , \ldots , L \} $.
Here $ ( x ) _ { 1 } $ denotes the first coordinate of $ x $.
For S with $ \Wid S < L / 2 $, a site $ \xl \in S $ is said to be the {\em left-most site} of $ S $ if $ ( \xl ) _ { 1 } - a = 0 $.
Similarly, a site $ \xr \in S $ is said to be the {\em right-most site} of $ S $ if $ ( \xr ) _ { 1 } - a = w - 1$.

We consider a fermion system on the lattice $ \Lambda $.
For $ x \in \L $ and $ \s = \up , \down $, we denote the {\em creation operator}, {\em annihilation operator}, and {\em number operator} of the fermion at site $ x $ with spin $ \s $ by $ \cp _ { x , \s }$, $\cm _ { x , \s }$, and $\n _ { x , \s } = \cp _ { x , \s } \cm _ { x , \s } $, respectively.
Note that $ ( \cp _ { x , \s } ) ^ { \dagger } = \cm _ { x , \s } $, $ ( \cm _ { x , \s } ) ^ { \dagger } = \cp _ { x , \s } $, and $ \n _ { x , \s } ^ { \dagger } = \n _ { x , \s } $.
The fermion operators satisfy the {\em anticommutation relations}
\begin{align}
	\left \{ \cp _ { x , \s } , \cm _ { y , \tau } \right \} = \delta _ { x , y } \delta _ { \s , \tau } , \label{anticommutation1}
\end{align}
and
\begin{align}
	\left \{ \cp _ { x , \s } , \cp _ { y , \tau } \right \} = \left \{ \cm _ { x , \s } , \cm _ { y , \tau } \right \} = 0 , \label{anticommutation2}
\end{align}
for any $ x , y \in \L $ and $ \s , \t = \up , \down $, where $ \{ \hA , \hB \} := \hA \hB + \hB \hA $. Throughout the present paper, we express creation/annihilation operators as $ \c _ { x , \s } ^ { \a } $ with $ \a = \pm $, $ \s = \up , \down $ and $ x \in \L $. We also use the shorthand notations $ \ao $ and $ \so $, defined by
\begin{align}
	\ao = - \a , \qquad \overline{ \up } = \down , \quad \text{and} \quad \overline{ \down } = \up .
\end{align}

We study the standard Hubbard model, whose Hamiltonian is
\begin{align}
	\hH = \Hhop + \Hint , \label{Hamiltonian}
\end{align}
with
\begin{align}
	\Hhop &= - t \sum _ { \substack{ x , y \in \Lambda \\ ( | x - y | = 1 ) } } \sum _ { \s = \up , \down } \cp _ { x , \s } \cm _ { y , \s } , \label{hopping term} \\
	\Hint &= U \sum _ { x \in \Lambda } \n _ { x , \up } \n _ { x , \down } \label{interaction term} ,
\end{align}
where $t\in\mathbb{R}$ is the hopping amplitude between adjacent sites, and $U\in\mathbb{R}$ represents the on-site (Coulomb) interaction.
Throughout the present paper, we assume $t\ne0$ and $U\ne0$.

By a {\em product} of fermion operators (which we mainly refer to as simply a product), we mean a finite product of $ \c _ { x , \s } ^ { \alpha } $ with $ \alpha = \pm $, $ \s = \up , \down $ and $ x \in \Lambda $. We always assume that the products are taken according to a suitable fixed ordering.
In this paper, we adopt the following convention for the ordering of operators\footnote{
It should be noted that this ordering applies to the basis for operator expansions, namely, the elements of $ \calPL $.
The terms constituting the Hamiltonian do not necessarily follow this convention.
} 
in a product:
Operators are arranged from left to right primarily in ascending order of the first coordinate component of the lattice site on which they act.
If the first components are identical, they are subsequently ordered by the second component, also in ascending order.
For operators acting on the same site, the operator acting on the up-spin is placed to the left of the one acting on the down-spin.
Finally, if both the site and the spin are the same, the ordering is determined by their type in the sequence of creation operator $ \cp $, annihilation operator $ \cm $, and number operator $ \n $.
Examples of such orderings are given below:
\eq
\cm _ { ( 1 , 4 , \ldots ) , \down } \cm _ { ( 2 , 3 , \ldots ) , \up } , \qquad
\cp _ { ( 1 , 1 , 1 , \ldots ) , \down } \n _ { ( 1 , 2 , 3 , \ldots ) , \up } \cm _ { ( 1 , 2 , 4 , \ldots ) , \up } , \qquad
\cm _ { x , \up } \cp _ { x , \down } \n _ { x , \down } , \qquad
\cm _ { y , \up } \n _ { y , \up } \cp _ { y , \down } \n _ { y , \down }
\en
By $ \calPL $ we denote the set of all products. The {\em support}, $ \Supp \sfA \subset \L $ of $ \sfA \in \calPL $ is a collection of sites on which $ \sfA $ acts in a nontrivial manner.

Note that the elements of $ \calPL $, with the identity $ \ho $, span the whole space of operators of the fermion system on $ \L $.
We define the widths of a product $ \sfA $ by
\begin{align}
	\Wid \sfA = \Wid \Supp \sfA .
\end{align}

We are almost ready to state our theorem.
Fix a constant $ \kb $ such that
\begin{align}
	1 \leq \kb \leq \frac{ L }{ 2 } .
\end{align}
We write the candidate of a local conserved quantity as
\begin{align}
	\hQ = \sum _ { \substack{ \sfA \in \calP _ { \L } \\ ( \Wid \sfA \leq \kb ) } } \qA \sfA , \label{Expansion of conserved quantities by products}
\end{align}
where $ \qA \in \mathbb{ C } $ are coefficients.
We further demand that there exists at least one product $ \sfA \in \calPL $ with $ \Wid \sfA = \kb $ such that $ \qA \neq 0 $.
We do not assume any symmetry, such as translational symmetry, for the coefficients $ \qA $.
This means that the candidate of a conserved quantity $ \hQ $ is a linear combination of products with the maximal width $ \kb $.

We say that $ \hQ $ is a local conserved quantity if and only if
\begin{align}
	[ \hQ , \hH ] = 0 . \label{definition of local conserved quantity}
\end{align}
Let us note here that one can assume $ \hQ $ is hermitian.
To see this it suffices to note $ [ \hQ _ { 0 } , \hH ] = 0 $ for any $ \hQ _ { 0 } $ implies $ [ \hQ _ { 0 } ^ { \dagger } , \hH ] = 0 $ and that $ \hQ _ { 0 } + \hQ _ { 0 } ^ { \dagger } $ and $ \i ( \hQ _ { 0 } - \hQ _ { 0 } ^ { \dagger } ) $ are hermitian.

Then, the following theorem is the main conclusion of the present paper.

\begin{theorem} \label{main theorem}
	There are no local conserved quantities $ \hQ $ with $ 3 \leq \kb \leq L / 2 $.
\end{theorem}

Note that the theorem is optimal since $ \hH ^ { 2 } $ is a conserved quantity with $ \kb = \frac{ L }{ 2 } + 2 $ when $ L $ is even.

Of course, the Hamiltonian $ \hH $ is a local conserved quantity with $ \kb = 2 $.
There are also local conserved quantities with $ \kb = 1 $ associated with the global spin-rotation and $ \eta $-pairing symmetries \cite{Yang1989}. In fact, we believe that one can also prove that these are the only conserved quantities with $\kb\leq 2 $.
See \cite{Fukai2023, Fukai2024} for a closely related result for the one-dimensional Hubbard model.

\section{Proof}

\subsection{Basic strategy and notation}
The proof here is based on the original strategy of Shiraishi \cite{Shiraishi2019, Shiraishi2024}, and follows the method developed by Shiraishi and Tasaki to treat the $d$-dimensional $S =\tfrac12$ XY and XYZ models \cite{ShiraishiTasaki2024}. 
Our proof, however, is not a straightforward extension of that in \cite{ShiraishiTasaki2024}.
There is an essential difficulty intrinsic to the Hubbard model, and our proof requires an extra step.
See the end of Section~\ref{s:strategy} and also Section~\ref{s:discussion}.

\subsubsection{Strategy of proof}
\label{s:strategy}

For a product $ \sfA \in \calPL $, the commutator with the Hamiltonian can be expressed as a linear combination of products as
\begin{align}
[ \sfA , \hH ] = \sum _ { \sfB \in \calPL } \lambda _ { \sfA , \sfB } \sfB . \label{Expanding the product}
\end{align}
The coefficients $ \lambda _ { \sfA , \sfB } $ are determined by the Hamiltonian \eqref{Hamiltonian} and the basic commutation relations \eqref{ccdc+}--\eqref{commutation3} of the fermionic operators. When $ \lambda _ { \sfA , \sfB } \neq 0 $, we say that $ \sfA $ {\em generates} $ \sfB $. We write the commutator between a general $ \hQ $ of the form \eqref{Expansion of conserved quantities by products} and $ \hH $ as
\begin{align}
	[ \hQ , \hH ] = \sum _ { \sfB \in \calPL } r _ { \sfB } \sfB ,
\end{align}
where the coefficient for $ \sfB $ is given by
\begin{align}
	r _ { \sfB } = \sum _ { \substack { \sfA \in \calPL \\ ( \Wid \sfA \leq \kb ) } } \lambda _ { \sfA , \sfB } q _ { \sfA } . \label{coefficient}
\end{align}
Since the products in $ \calPL $ are linearly independent, we see that the condition \eqref{definition of local conserved quantity} for a conserved quantity is equivalent to
\begin{align}
	r _ { \sfB } = 0 , \label{r=0}
\end{align}
for all $ \sfB \in \calPL $.

We regard \eqref{r=0} for all $\sfB\in\calPL$ with \eqref{coefficient}  as coupled linear equations for determining the unknown coefficients $q_{\sfA}$.
For $3\le\kb\le L/2$, by analyzing \eqref{r=0} for slected products $\sfB$, we shall prove that $q_{\sfA}=0$ for any $\sfA\in\calPL$ such that $\Wid\sfA=\kb$.
This contradicts the assumption that there is $\sfA$ with $\Wid\sfA=\kb$ and $q_{\sfA}\ne0$, and hence proves Theorem~\ref{main theorem}.

Let us note here that there is an essential difference between the proof in \cite{ShiraishiTasaki2024} for quantum spin systems and the current proof for the Hubbard model.
In \cite{ShiraishiTasaki2024} (and in many, but not all, similar works for quantum spin systems), it is enough to consider the relation \eqref{r=0} for some products $\sfB$ with $\Wid\sfB=\kb+1$ and $\Wid\sfB=\kb$.
In the present work, on the other hand, it is necessary to consider \eqref{r=0} for some $\sfB$ with $\Wid\sfB=\kb+1$, $\Wid\sfB=\kb$, and $\Wid\sfB=\kb-1$.
In other words, the proof in \cite{ShiraishiTasaki2024} consists of two steps, while that in the present work consists of three steps.
This reflects the essential difficulty encountered in the Hubbard model.
See footnote~\ref{fn:Shiraishi} and Section~\ref{s:discussion}.

As a special case of \eqref{coefficient} and \eqref{r=0}, we obtain the following lemma. This lemma is repeatedly used throughout this paper.

\begin{lemma}\label{one product generate relation}
Let $ \sfB \in \calPL $ be a product that is generated by a unique product $ \sfA \in \calPL $ with $ \Wid \sfA \leq \kb $ (i.e., $ \lambda _ { \sfA , \sfB } \neq 0 $ and $ \lambda _ { \sfA ^ { \p } , \sfB } = 0 $ for all other $ \sfA ^ { \p } \in \calPL \setminus \{ \sfA \} $ with $ \Wid \sfA ^ { \p } \leq \kb $). Then, we have $ \qA = 0 $.
\end{lemma}

\subsubsection{Basic commutation relations}
We need to evaluate the commutator $ [ \sfA , \hH ] $ for various $ \sfA \in \calPL $. From the anticommutation relations of the fermionic operators \eqref{anticommutation1}, \eqref{anticommutation2}, we obtain the following {\em commutation relations}. For any $ x , y \in \Lambda $ with $ x \neq y $ and $ \s = \up , \down $, we have
\begin{align}
	\Big [ \cm _ { x , \s } \ , \ \cp _ { x , \s } \cm _ { y , \s } \Big ] &= + \cm _ { y , \s } , \label{ccdc+} \\
	\Big [ \cp _ { x , \s } \ , \ \cp _ { y , \s } \cm _ { x , \s } \Big ] &= - \cp _ { y , \s } , \label{cdcdc-} \\
	\Big [ \cm _ { x , \s } \ , \ \n _ { x , \up } \n _ { x , \down } \Big ] &= + \cm _ { x , \s } \n _ { x , \so } , \label{cnn+} \\
	\Big [ \cp _ { x , \s } \ , \ \n _ { x , \up } \n _ { x , \down } \Big ] &= - \cp _ { x , \s } \n _ { x , \so } , \label{cdnn-} \\
	\Big [ \n _ { x , \s } \ , \ \cp _ { x , \s } \cm _ { y , \s } \Big ] &= + \cp _ { x , \s } \cm _ { y , \s } , \label{ncdc+} \\
	\Big [ \n _ { x , \s } \ , \ \cp _ { y , \s } \cm _ { x , \s } \Big ] &= - \cp _ { y , \s } \cm _ { x , \s } . \label{ncdc-}
\end{align}
Recall that $ \overline{ \up } = \down $ and $ \overline{ \down } = \up $. Note that the products appearing in the second slot of the commutators are all parts of the Hamiltonian. Organizing these commutation relations using $ \a = \pm $, we have
\begin{align}
	\Big [ \c _ { x , \s } ^ { \a } \ , \ \c _ { x , \s } ^ { \ao } \c _ { y , \s } ^ { \a } \Big ] &= \c _ { y , \s } ^ { \a } , \label{commutation1} \\
	\Big [ \c _ { x , \s } ^ { \a } \ , \ \n _ { x , \up } \n _ { x , \down } \Big ] &= \ao \c _ { x , \s } ^ { \a } \n _ { x , \so } , \label{commutation2} \\
	\Big [ \n _ { x , \s } \ , \ \c _ { x , \s } ^ { \a } \c _ { y , \s } ^ { \ao } \Big ] &= \a \c _ { x , \s } ^ { \a } \c _ { y , \s } ^ { \ao } .\label{commutation3}
\end{align}
Again recall that $ \ao = - \a $. These commutation relations are frequently used throughout this paper.

In the present paper, we employ a graphical representation as a method to clearly grasp the structure of complex products of fermionic operators.
For simplicity, we illustrate them on a ladder represented by $ \{ 1 , 2 , \ldots , L \} \times \{ 1 , 2 \} $, but this does not entail any loss of generality.
For general $ d \geq 2 $, a coordinate such as $ (1, 2) $ is interpreted as an abbreviation of $ (1, 2, 0, \ldots, 0) $.
Here, the unit vector in the first direction is defined as
\begin{align}
\ei = ( 1 , 0 , \ldots , 0 ).
\end{align}

We represent the fermionic creation, annihilation, and number operators by
\tikz{ \fill (1, 1) circle (1.4mm) ; } $ = \cp $, 
\tikz{ \node[draw, circle, inner sep=1mm, line width=0.5mm] at (3, 1) {} ; } $ = \cm $, and
\tikz{ \node[draw, inner sep=1.2mm, line width=0.5mm] at (0, 0) {} ; } $ = \n $.
Using this, the commutation relation
\eq
\Big [ \ \cp _ { ( 1 , 1 ) , \up } \cm _ { ( 2 , 1 ) , \down } \cp _ { ( 2 , 2 ) , \down } \cm _ { ( 3 , 1 ) , \up } \n _ { ( 3 , 1 ) ,\down } \ , \ \cp _ { ( 3 , 1 ) , \up } \cm _ { ( 4 , 1 ) , \up } \ \Big ] \ 
= \ \cp _ { ( 1 , 1 ) , \up } \cm _ { ( 2 , 1 ) , \down } \cp _ { ( 2 , 2 ) , \down } \n _ { ( 3 , 1 ) ,\down } \cm _ { ( 4 , 1 ) , \up } ,
\en
for example, can be illustrated as follows:
\begin{align}
	\left [ \ 
\begin{tikzpicture}[tdplot_main_coords, scale=0.7, baseline=2.5mm]
	% ボンド	
	\foreach \x in {1, 2, ..., 5} {
	\foreach \y in {1, 2} {
	\foreach \z in {0, 1} {
		\draw (\x, \y, 0) -- (\x, \y, 1) ;
		\draw (0.75, \y, \z) -- (5.25, \y, \z) ;
		\draw (\x, 0.75, \z) -- (\x, 2.25, \z) ;
	} } }
	% 横
	\node at (1, 1, -0.5) {1} ;
	\node at (2, 1, -0.5) {2} ;
	\node at (3, 1, -0.5) {3} ;
	\node at (4, 1, -0.5) {4} ;
	\node at (5, 1, -0.5) {5} ;
	% 奥行
	\node at (0.5, 2, 1) {1} ;
	\node at (0.5, 1, 1) {2} ;
	% 縦
	\node at (5.5, 2, 1) {$\up$} ;
	\node at (5.5, 2, 0) {$\down$} ;
	% 生成演算子
	\node[fill, circle] at (1, 2, 1) {} ;
	\node[fill, circle] at (2, 1, 0) {} ;
	% 消滅演算子
	\node[fill=white, draw, circle, inner sep=1mm, line width=0.5mm] at (3, 2, 1) {} ;
	\node[fill=white, draw, circle, inner sep=1mm, line width=0.5mm] at (2, 2, 0) {} ;
	% 個数演算子
	\node[fill=white, draw, inner sep=1.2mm, line width=0.5mm] at (3, 2, 0) {} ;
\end{tikzpicture} \ , \ 
\begin{tikzpicture}[tdplot_main_coords, scale=0.7, baseline=2.5mm]
	% ボンド	
	\foreach \x in {1, 2, ..., 5} {
	\foreach \y in {1, 2} {
	\foreach \z in {0, 1} {
		\draw (\x, \y, 0) -- (\x, \y, 1) ;
		\draw (0.75, \y, \z) -- (5.25, \y, \z) ;
		\draw (\x, 0.75, \z) -- (\x, 2.25, \z) ;
	} } }
	% 横
	\node at (1, 1, -0.5) {1} ;
	\node at (2, 1, -0.5) {2} ;
	\node at (3, 1, -0.5) {3} ;
	\node at (4, 1, -0.5) {4} ;
	\node at (5, 1, -0.5) {5} ;
	% 奥行
	\node at (0.5, 2, 1) {1} ;
	\node at (0.5, 1, 1) {2} ;
	% 縦
	\node at (5.5, 2, 1) {$\up$} ;
	\node at (5.5, 2, 0) {$\down$} ;
	% 生成演算子
	\node[fill, circle] at (3, 2, 1) {} ;
	% 消滅演算子
	\node[fill=white, draw, circle, inner sep=1mm, line width=0.5mm] at (4, 2, 1) {} ;
\end{tikzpicture} \ \right ] \ &= \ 
\begin{tikzpicture}[tdplot_main_coords, scale=0.7, baseline=2.5mm]
	% ボンド	
	\foreach \x in {1, 2, ..., 5} {
	\foreach \y in {1, 2} {
	\foreach \z in {0, 1} {
		\draw (\x, \y, 0) -- (\x, \y, 1) ;
		\draw (0.75, \y, \z) -- (5.25, \y, \z) ;
		\draw (\x, 0.75, \z) -- (\x, 2.25, \z) ;
	} } }
	% 横
	\node at (1, 1, -0.5) {1} ;
	\node at (2, 1, -0.5) {2} ;
	\node at (3, 1, -0.5) {3} ;
	\node at (4, 1, -0.5) {4} ;
	\node at (5, 1, -0.5) {5} ;
	% 奥行
	\node at (0.5, 2, 1) {1} ;
	\node at (0.5, 1, 1) {2} ;
	% 縦
	\node at (5.5, 2, 1) {$\up$} ;
	\node at (5.5, 2, 0) {$\down$} ;
	% 生成演算子
	\node[fill, circle] at (1, 2, 1) {} ;
	\node[fill, circle] at (2, 1, 0) {} ;
	% 消滅演算子
	\node[fill=white, draw, circle, inner sep=1mm, line width=0.5mm] at (4, 2, 1) {} ;
	\node[fill=white, draw, circle, inner sep=1mm, line width=0.5mm] at (2, 2, 0) {} ;
	% 個数演算子
	\node[fill=white, draw, inner sep=1.2mm, line width=0.5mm] at (3, 2, 0) {} ;
\end{tikzpicture} \label{figure}
\end{align}
To explain the notation in \eqref{figure} in more detail, the horizontal direction corresponds to the spatial axis of the first direction. The positive direction is to the right.
The depth direction corresponds to the spatial axis of the second direction (for the proof, this does not necessarily have to be the second direction; any axis different from the first direction suffices). The positive direction is toward the front.
The vertical direction represents the spin. The upper layer corresponds to $ \up $, and the lower layer corresponds to $ \down $.
Here, the products illustrated are also assumed to follow the operator ordering defined in Section~\ref{s:def}.

\subsection{First step: basic relations for products with width $ \kb + 1 $}

In the present subsection (and only in the present subsection), we express any product $ \sfA \in \calPL $ in the form
\begin{align}
	\sfA = \pm \prod _ { x \in \Supp \sfA } \hA _ { x }
\end{align}
where the local operator $ \hA _ { x } $ is either $ \c _ { x , \s } ^ { \alpha } $, $ \n _ { x , \s } $, $ \c _ { x , \up } ^ { \alpha } \c _ { x , \down } ^ { \beta } $, $ \c _ { x , \s } ^ { \alpha } \n _ { x , \so } $, or $ \n _ { x , \up } \n _ { x , \down } $ with $ \alpha , \beta = \pm $, $ \s = \up , \down $.

The following lemma and its proof represent an essential idea used in the proof in the present work.

\begin{lemma}\label{leftright-most site conditions}
	For $\kb$ with $ 2 \leq \kb \leq L / 2 $, let $ \sfA \in \calPL $ be such that $ \Wid \sfA = \kb $. One has $ \qA = 0 $ unless both the following two conditions are satisfied:
	\begin{enumerate}[(i)]
		\item $ \Supp \sfA $ has a unique left-most site $ \xl $ with $ \hA _ { \xl } = \c _ { \xl , \s } ^ { \a } $ with some $ \a = \pm $ and $ \s = \up , \down $. It also holds that $ \hA _ { \xl + \ei } = \c _ { \xl + \ei , \s } ^ { \ao } $ or $ \xl + \ei \not \in \Supp \sfA $.
		\item $ \Supp \sfA $ has a unique right-most site $ \xr $ with $ \hA _ { \xr } = \c _ { \xr , \t } ^ { \b } $ with some $ \b = \pm $ and $ \t = \up , \down $. It also holds that $ \hA _ { \xr - \ei } = \c _ { \xr - \ei , \t } ^ { \bo } $ or $ \xr - \ei \not \in \Supp \sfA $.
	\end{enumerate}
\end{lemma}
{\it Proof:} Let $ \xr $ be a right-most site of $ \sfA $. We define a product $ \sfB \in \calPL $ by
\begin{align}
	\sfB = \pm [ \sfA , \c _ { \xr , \s } ^ { + } \c _ { \xr + \ei , \s } ^ { - } ] \quad \text{ or } \quad \sfB = \pm [ \sfA , \c _ { \xr + \ei , \s } ^ { + } \c _ { \xr , \s } ^ { - } ] . \label{definition of commutator manner}
\end{align}
Examining the commutation relations \eqref{ccdc+}--\eqref{commutation3}, one finds that at least one of them for some $ \s = \up , \down $ is nonzero. Note that the commutator adds a new site $ \xr + \ei $ to the support of $ \sfA $, so that $ \Wid \sfB = \kb + 1 $. By definition, $ \sfA $ generates $ \sfB $.
Next, consider whether there exists another product $ \sfA ^ { \p } $ with $ \Wid \sfA ^ { \p } \leq \kb $ that also generates $ \sfB $. If no such $ \sfA ^ { \p } $ exists, then $ \sfA $ is the unique product generating $ \sfB $, and by Lemma 3.1, we have $ \qA = 0 $.

Note that any other product $ \sfA ^ { \p } $ (different from $ \sfA $) with $ \Wid \sfA ^ { \p } \leq \kb $ generating $ \sfB $ must satisfy $ \Supp \sfA ^ { \p } = \Supp \sfB \setminus \{ \xl \} $, where $ \xl $ is the unique left-most site of $ \sfB $ (and therefore also the unique left-most site of $ \sfA $).
Also, for $ \t = \up , \down $, either $ [ \sfA ^ { \p } , \cp _ { \xl , \t } \cm _ { \xl + \ei , \t } ] $ or $ [ \sfA ^ { \prime } , \cp _ { \xl + \ei , \t } \cm _ { \xl , \t } ] $ must be proportional to $ \sfB $. This implies that condition (i) holds. We have shown that condition (i) is necessary for $ \qA $ to be nonzero.

By swapping the right-most and left-most sites and repeating the same argument, it is also found that condition (ii) is necessary for $ \qA $ to be nonzero.
$ \blacksquare $\\

The above proof contains the essential idea of the procedure called the Shiraishi shift.

To get the idea, let $ \kb \geq 3 $, and suppose that $ \sfA \in \calPL $ with $ \Wid \sfA = \kb $ satisfies conditions (i) and (ii) of Lemma 3.2.

Consider, e.g., the case with $ \hA _ { \xl } = \cm _ { \xl , \s } $, $ \hA _ { \xl + \ei } = \cp _ { \xl + \ei , \s } $, and $ \hA _ { \xr } = \cm _ { \xr , \t } $ for some $ \s , \t = \up , \down $.
See Figure \ref{idea of Shiraishi shift}.
\begin{figure}[htb]
\center
\begin{minipage}{0.3\linewidth}
\center
$ \sfA = $
\begin{tikzpicture}[tdplot_main_coords, scale=0.7, baseline=2.5mm]
	% ボンド	
	\foreach \x in {1, 2, ..., 5} {
	\foreach \y in {1, 2} {
	\foreach \z in {0, 1} {
		\draw (\x, \y, 0) -- (\x, \y, 1) ;
		\draw (0.75, \y, \z) -- (5.25, \y, \z) ;
		\draw (\x, 0.75, \z) -- (\x, 2.25, \z) ;
	} } }
	% 生成演算子
	\node[fill, circle] at (3, 1, 0) {} ;
	\node[fill, circle] at (2, 1, 0) {} ;
	% 消滅演算子
	\node[fill=white, draw, circle, inner sep=1mm, line width=0.5mm] at (1, 1, 0) {} ;
	\node[fill=white, draw, circle, inner sep=1mm, line width=0.5mm] at (4, 2, 1) {} ;
	% 個数演算子
	\node[fill=white, draw, inner sep=1.2mm, line width=0.5mm] at (3, 2, 0) {} ;
\end{tikzpicture}
\end{minipage}
\quad
\begin{minipage}{0.3\linewidth}
\center
$ \sfB = $
\begin{tikzpicture}[tdplot_main_coords, scale=0.7, baseline=2.5mm]
	% ボンド	
	\foreach \x in {1, 2, ..., 5} {
	\foreach \y in {1, 2} {
	\foreach \z in {0, 1} {
		\draw (\x, \y, 0) -- (\x, \y, 1) ;
		\draw (0.75, \y, \z) -- (5.25, \y, \z) ;
		\draw (\x, 0.75, \z) -- (\x, 2.25, \z) ;
	} } }
	% 生成演算子
	\node[fill, circle] at (3, 1, 0) {} ;
	\node[fill, circle] at (2, 1, 0) {} ;
	% 消滅演算子
	\node[fill=white, draw, circle, inner sep=1mm, line width=0.5mm] at (1, 1, 0) {} ;
	\node[fill=white, draw, circle, inner sep=1mm, line width=0.5mm] at (5, 2, 1) {} ;
	% 個数演算子
	\node[fill=white, draw, inner sep=1.2mm, line width=0.5mm] at (3, 2, 0) {} ;
\end{tikzpicture}
\end{minipage}
\quad
\begin{minipage}{0.33\linewidth}
\center
$ \sfA ^ { \p } = \calS ( \sfA ) = $
\begin{tikzpicture}[tdplot_main_coords, scale=0.7, baseline=2.5mm]
	% ボンド	
	\foreach \x in {1, 2, ..., 5} {
	\foreach \y in {1, 2} {
	\foreach \z in {0, 1} {
		\draw (\x, \y, 0) -- (\x, \y, 1) ;
		\draw (0.75, \y, \z) -- (5.25, \y, \z) ;
		\draw (\x, 0.75, \z) -- (\x, 2.25, \z) ;
	} } }
	% 生成演算子
	\node[fill, circle] at (3, 1, 0) {} ;
	% 消滅演算子
	\node[fill=white, draw, circle, inner sep=1mm, line width=0.5mm] at (5, 2, 1) {} ;
	% 個数演算子
	\node[fill=white, draw, inner sep=1.2mm, line width=0.5mm] at (2, 1, 0) {} ;
	\node[fill=white, draw, inner sep=1.2mm, line width=0.5mm] at (3, 2, 0) {} ;
\end{tikzpicture}
\end{minipage}
\captionsetup{width=0.8\linewidth}
\caption[dummy]{An example of the Shiraishi shift for $ \kb = 4 $.
Here, $ \sfA = \cm _ { ( 1 , 2 ) , \down } \cp _ { ( 2 , 2 ) , \down } \n _ { ( 3 , 1 ) , \down } \cp _ { ( 3 , 2 ) , \down } \cm _ { ( 4 , 1 ) , \up } $ is a product of width $ \Wid \sfA = 4 $ that satisfies conditions (i) and (ii) of Lemma 3.2.
The product $ \sfB = \cm _ { ( 1 , 2 ) , \down } \cp _ { ( 2 , 2 ) , \down } \n _ { ( 3 , 1 ) , \down } \cp _ { ( 3 , 2 ) , \down } \cm _ { ( 5 , 1 ) , \up } $ is defined from the commutator \eqref{definition of sfB from cimmutator}.
Since $ \cm $ is added to the right-most site, the width becomes $ \Wid \sfB = 5 $.
Next, by removing $ \cm $ at the left-most site, we obtain $ \sfA ^ { \p } $ such that the commutation relation \eqref{Shift fulfillment relationship} holds.
The resulting product $ \sfA ^ { \p } $ with $ \Wid \sfA ^ { \p } = 4 $ is the Shiraishi shift $ \calS ( \sfA ) $.
Here, $ \Supp \calS ( \sfA ) $ has a unique left-most site, but condition (i) is not satisfied because $ \hA ^ { \p } _ { \xl } \neq \c _ { \xl } ^ { \pm } $.
Therefore, $ \calS ^ { 2 } ( \sfA ) $ does not exist, and $ q _ { \calS ( \sfA ) } = 0 $.
Hence, from \eqref{E:proportional relation}, it follows that $ \qA = 0 $.
}
\label{idea of Shiraishi shift}
\end{figure}
As in \eqref{definition of commutator manner}, we define $ \sfB \in \calPL $ by
\begin{align}
	\sfB &= \pm [ \sfA , \cp _ { \xr , \t } \cm _ { \xr + \ei , \t } ] \notag \\
	&= \pm \Big ( \prod _ { y \in \Supp \sfA \setminus \{ \xr \} } \hA _ { y } \Big ) [ \cm _ { \xr , \t } , \cp _ { \xr , \t } \cm _ { \xr + \ei , \t } ] \notag \\
	&= \pm \Big ( \prod _ { y \in \Supp \sfA \setminus \{ \xr \} } \hA _ { y } \Big ) \cm _ { \xr + \ei , \t } , \label{definition of sfB from cimmutator}
\end{align}
where the $ \pm $ signs are not taken consistently. We used the commutation relation \eqref{ccdc+}.
Note that $ \Wid \sfB = \kb + 1 $.
Furthermore, since $ \hB _ { \xl } = \cm _ { \xl , \s } $ and $ \hB _ { \xl + \ei } = \cp _ { \xl + \ei , \s } $,
\begin{align}
	\sfA ^ { \p } = \pm \Big ( \prod _ { y \in \Supp \sfB \setminus \{ \xl , \xl + \ei \} } \hB _ { y } \Big ) \n _ { \xl + \ei , \s } ,
\end{align}
generates $ \sfB $ as
\begin{align}
	\sfB &= \pm [ \sfA ^ { \p } , \cp _ { \xl + \ei , \s } \cm _ { \xl , \s } ] \notag \\
	&= \pm \Big ( \prod _ { y \in \Supp \sfA ^ { \p } \setminus \{ \xl + \ei \} } \hA ^ { \prime } _ { y } \Big ) [ \n _ { \xl + \ei , \s } , \cp _ { \xl + \ei , \s } \cm _ { \xl , \s } ] \notag \\
	&= \pm \Big ( \prod _ { y \in \Supp \sfA ^ { \p } \setminus \{ \xl + \ei \} } \hA ^ { \p } _ { y } \Big ) \cp _ { \xl + \ei , \s } \cm _ { \xl , \s } . \label{Shift fulfillment relationship}
\end{align}
We used the commutation relation \eqref{ncdc+}. Note that, $ \Wid \sfA ^ { \p } = \kb $.
Clearly $ \sfA $ and $ \sfA ^ { \p } $ are the only products with width $ \leq \kb $ that generate $ \sfB $.
From \eqref{Hamiltonian}, \eqref{definition of sfB from cimmutator}, and \eqref{Shift fulfillment relationship}, the coefficients in the expansion \eqref{Expanding the product} are determined as $ \lambda _ { \sfA , \sfB } = - t $ and $ \lambda _ { \sfA^\p , \sfB } = - t $.
Therefore, the coefficient \eqref{coefficient} for $ \sfB $ is given by
\begin{align}
	r _ { \sfB } = - t q _ { \sfA } - t q _ { \sfA ^ { \p } } \label{example coefficient}
\end{align}
By requiring $ r _ { \sfB } = 0 $, we find $ q _ { \sfA ^ { \p } } = - q _ { \sfA } $.
We denote $ \sfA ^ { \p } $ by $ \calS ( \sfA ) $ and call it the {\em Shiraishi shift} of $ \sfA $.

This procedure can be generalized to define the Shiraishi shift $ \calS ( \sfA ) \in \calPL $ for $ \sfA \in \calPL $ with $ \Wid \sfA = \kb $.
Here we consider general $ \kb $ with $ 2 \leq \kb \leq L / 2 $.
If $ \sfA $ does not satisfy conditions (i) or (ii) of Lemma 3.2, we say the Shiraishi shift does not exist.
In this case, from Lemma 3.2, it follows that $ \qA = 0 $.
If $ \sfA $ satisfies (i) and (ii), we define $ \sfB \in \calPL $ by
\begin{align}
	\sfB :=
	\begin{dcases*}
		\pm [ \sfA , \cp _ { \xr , \s } \cm _ { \xr + \ei , \s } ] & if $ \hA _ { \xr } = \cm _ { \xr , \s } $; \\
		\pm [ \sfA , \cp _ { \xr + \ei , \s } \cm _ { \xr , \s } ] & if $ \hA _ { \xr } = \cp _ { \xr , \s } $.
	\end{dcases*}
\end{align}
We then let $ \sfA ^ { \p } \in \calPL $ be a product that satisfy
\begin{align}
	\sfB =
	\begin{dcases*}
		\pm [ \sfA ^ { \p } , \cp _ { \xl + \ei , \t } \cm _ { \xl , \t } ] & if $ \hA _ { \xl } = \cm _ { \xl , \t } $; \\
		\pm [ \sfA ^ { \p } , \cp _ { \xl , \t } \cm _ { \xl + \ei , \t } ] & if $ \hA _ { \xl } = \cp _ { \xl , \t } $.
\end{dcases*} \label{definition of shift from commutator}
\end{align}
If there is such $ \sfA ^ { \p } $, then we denote it as $ \calS ( \sfA ) $.
If no such $ \sfA ^ { \p } $ exists, then we say that $ \calS ( \sfA ) $ does not exist.
In this case, the only product with width $ \leq \kb $ generating $ \sfB $ is $ \sfA $, and from Lemma 3.1, it follows that $ \qA = 0 $.
When the shift $ \calS ( \sfA ) $ exists, the coefficients $ \qA $ and $ q _ { \calS ( \sfA ) } $ are related as in \eqref{example coefficient}.

We summarize these observations as the following lemma.

\begin{lemma} \label{L:coefficient of Shift}
	For $\kb$ with $ 2 \leq \kb \leq L / 2 $, let $ \sfA \in \calPL $ be such that $ \Wid \sfA = \kb $. We have $ \qA = 0 $ if $ \calS ( \sfA ) $ does not exist. If $ \calS ( \sfA ) $ exist, we have
	\begin{align}
		q _ { \calS ( \sfA ) } = \pm \qA . \label{E:proportional relation}
	\end{align}
\end{lemma}

By applying the Shiraishi shift repeatedly, we can further restrict the form of products with possibly nonzero coefficients.

The following lemma is the main result in the present subsection.

\begin{lemma}\label{L:two site product}
	For $\kb$ with $ 2 \leq \kb \leq L / 2 $, let $ \sfA \in \calPL $ be such that $ \Wid \sfA = \kb $. One has $ \qA = 0 $ unless
	\begin{align}
		\sfA = \c _ { x , \s } ^ { \alpha } \c _ { y , \t } ^ { \beta } , \label{two site product}
	\end{align}
	for some $ \alpha , \beta = \pm $, $ \s , \t = \up , \down $, and $ x , y \in \L $ such that $ \big ( y - x \big ) _ { 1 } = \kb - 1 $. Furthermore for $ \sfA $ as in \eqref{two site product} we have
	\begin{align}
		q _ { \calS ( \sfA ) } = - \alpha \beta \qA , \label{E:coefficient of Shift}
	\end{align}
	where the Shiraishi shift of $ \sfA $ is
	\begin{align}
		\calS ( \sfA ) = \c _ { x + \ei , \s } ^ { \alpha } \c _ { y + \ei , \t } ^ { \beta }.
	\end{align}
\end{lemma}
We here assumed that the sign convention for the set $ \calPL $ of products is chosen so that $ \sfA , \calS ( \sfA ) \in \calPL $.

Let us note here that Lemma 3.4 is the most we can get from the relations $ r _ { \sfB } = 0 $ for $ \sfB \in \calPL $ with $ \Wid \sfB = \kb + 1 $. It is worth comparing the situation with the corresponding results for spin systems, namely, Lemma 3.6 of \cite{ShiraishiTasaki2024} and Lemma 3.4 of \cite{FutamiTasaki2025}, where possible products are restricted to essentially one-dimensional strings. In the present case of the Hubbard model, on the other hand, the relative location of two sites $ x $ and $ y $ is still quite arbitrary.\\

\noindent{\it Proof:}\/
For $ \kb = 2 $, we see from conditions (i) and (ii) that, a product with $ \Wid \sfA = 2 $ with possibly nonzero coefficient $ \qA $ takes the form $ \sfA = \c _ { x } ^ { \a } \c _ { y } ^ { \b } $ with $ \a , \b = \pm $ and $ \s , \t = \up , \down $, where $ \big ( y - x \big ) _ { 1 } = 1 $. This is exactly the form of \eqref{two site product}.

We shall treat the case with $ 3 \leq \kb \leq L / 2 $.
Let us assume that $ \sfA \in \calPL $ satisfies the conditions (i), (ii) of Lemma 3.2 and hence $ \sfA ^ { \p } = \calS ( \sfA ) $ exists.
We shall examine the necessary conditions for $ \sfA ^ { \p } $ to satisfy the condition (i). Since $ \xl + \ei $ is the left-most site of $ \sfA ^ { \p } $, the condition (i) for $ \sfA ^ { \p } $ requires $ \hA _ { \xl + \ei } ^ { \p } = \c _ { \xl + \ei , \zeta } ^ { \gamma } $ with $ \gamma = \pm $, $ \zeta = \up , \down $.
Recalling the construction \eqref{definition of shift from commutator} of $ \sfB $, one finds from the commutation relations \eqref{ccdc+} or \eqref{cdcdc-} that $ \xl + \ei \not \in \Supp \sfB $.
Since $ \kb \geq 3 $, this implies $ \xl + \ei \not \in \Supp \sfA $. Noting that $ \hA _ { \xl } = \c _ { \xl , \t } ^ { \b } $ from the condition (i), we see that the two left-most sites in $ \Supp \sfA $ precisely coincide with the desired form \eqref{two site product}.
We get the desired result by repeating this procedure for $ \kb - 1 $ times.

The relation \eqref{E:coefficient of Shift} for the coefficients follows by generalizing the expression \eqref{example coefficient}.
$ \blacksquare $\\

\subsection{Second step: basic relations for products with width $ \kb $}

In this subsection, we use the relations that generate products with width $ \kb $ to prove Lemma 3.5, Lemma 3.6, and Lemma 3.7, which determine the form of products with $ \Wid = \kb $. Recall that Lemma 3.4 shows the only relevant products with $ \Wid = \kb $ are of the form \eqref{two site product}. Here, we state the following lemma, which represents the effect of interaction terms appearing for the first time in this paper.

\begin{lemma}\label{L:horizontal condition}
For $\kb$ with $ 2 \leq \kb \leq L / 2 $, let $ \sfA $ be of the form \eqref{two site product} with arbitrary $ \alpha , \beta = \pm $, $ \s , \tau = \up , \down $, and $ x , y \in \L $ such that $ \big ( y - x \big ) _ { 1 } = \kb - 1 $.
Then we have $ \qA = 0 $ unless
\eq
y = x + ( \kb - 1 ) \ei . \label{horizontal site}
\en
\end{lemma}

The lemma states that the two ends of the product must be aligned horizontally, thus essentially reducing our problem to that in one dimension.

\medskip
\noindent{\it Proof:}\/
Fix arbitrary $ \alpha , \beta = \pm $, $ \s , \t = \up , \down $.
Take $ x , y \in \L $ such that $ \big ( y - x \big ) _ { 1 } = \kb - 1 $ and do not satisfy \eqref{horizontal site}.
We shall show $ \qA = 0 $.

For simplicity, we shall restrict ourselves to the case with $ d = 2$, and assume $ x = ( 1 , 1 ) $ without losing generality. Then our $ \sfA $ is
\begin{align*}
	\sfA = \c _ { ( 1 , 1 ) , \s } ^ { \alpha } \c _ { ( \kb , m ) , \t } ^ { \beta } ,
\end{align*}
with $ m \neq 1 $. We define
\begin{align}
	\sfD _ { j } ^ { \p } &= \c _ { ( j , 1 ) , \s } ^ { \a } \n _ { ( \kb , m ) , \to } \c _ { ( j + \kb - 1 , m ) , \t } ^ { \b } ,
\end{align}
for $ j = 1 , \ldots , \kb $, and
\begin{align}
	\sfE _ { j } ^ { \p } &= \c _ { ( j , 1 ) , \s } ^ { \a } \n _ { ( \kb , m ) , \to } \c _ { ( j + \kb - 2 , m ) , \t } ^ { \b } ,
\end{align}
for $ j = 2 , \ldots , \kb $. Note that $ \Wid \sfA = \Wid \sfD _ { j } ^ { \p } = \kb $ and $ \Wid \sfE _ { j } ^ { \p } = \kb - 1 $. See Figure \ref{F:horizontal condition}.

\begin{figure}[htb]
\center
\begin{minipage}{0.3\linewidth}
\center
$ \sfA = $
\begin{tikzpicture}[tdplot_main_coords, scale=0.7, baseline=2.5mm]
	% ボンド	
	\foreach \x in {1, 2, ..., 5} {
	\foreach \y in {1, 2} {
	\foreach \z in {0, 1} {
		\draw (\x, \y, 0) -- (\x, \y, 1) ;
		\draw (0.75, \y, \z) -- (5.25, \y, \z) ;
		\draw (\x, 0.75, \z) -- (\x, 2.25, \z) ;
	} } }
	% 生成演算子
	\node[fill, circle] at (1, 2, 1) {} ;
	% 消滅演算子
	\node[fill=white, draw, circle, inner sep=1mm, line width=0.5mm] at (3, 1, 0) {} ;
\end{tikzpicture}

\end{minipage}
\quad
\begin{minipage}{0.3\linewidth}
\center
$ \sfD _ { 1 } ^ { \p } = $
\begin{tikzpicture}[tdplot_main_coords, scale=0.7, baseline=2.5mm]
	% ボンド	
	\foreach \x in {1, 2, ..., 5} {
	\foreach \y in {1, 2} {
	\foreach \z in {0, 1} {
		\draw (\x, \y, 0) -- (\x, \y, 1) ;
		\draw (0.75, \y, \z) -- (5.25, \y, \z) ;
		\draw (\x, 0.75, \z) -- (\x, 2.25, \z) ;
	} } }
	% 生成演算子
	\node[fill, circle] at (1, 2, 1) {} ;
	% 消滅演算子
	\node[fill=white, draw, circle, inner sep=1mm, line width=0.5mm] at (3, 1, 0) {} ;
	% 個数演算子
	\node[fill=white, draw, inner sep=1.2mm, line width=0.5mm] at (3, 1, 1) {} ;
\end{tikzpicture}

\end{minipage}\\
\vspace{1mm}
\begin{minipage}{0.3\linewidth}
\center
$ \sfE _ { 2 } ^ { \p } = $
\begin{tikzpicture}[tdplot_main_coords, scale=0.7, baseline=2.5mm]
	% ボンド	
	\foreach \x in {1, 2, ..., 5} {
	\foreach \y in {1, 2} {
	\foreach \z in {0, 1} {
		\draw (\x, \y, 0) -- (\x, \y, 1) ;
		\draw (0.75, \y, \z) -- (5.25, \y, \z) ;
		\draw (\x, 0.75, \z) -- (\x, 2.25, \z) ;
	} } }
	% 生成演算子
	\node[fill, circle] at (2, 2, 1) {} ;
	% 消滅演算子
	\node[fill=white, draw, circle, inner sep=1mm, line width=0.5mm] at (3, 1, 0) {} ;
	% 個数演算子
	\node[fill=white, draw, inner sep=1.2mm, line width=0.5mm] at (3, 1, 1) {} ;
\end{tikzpicture}

\end{minipage}
\quad
\begin{minipage}{0.3\linewidth}
\center
$ \sfD _ { 2 } ^ { \p } = $
\begin{tikzpicture}[tdplot_main_coords, scale=0.7, baseline=2.5mm]
	% ボンド	
	\foreach \x in {1, 2, ..., 5} {
	\foreach \y in {1, 2} {
	\foreach \z in {0, 1} {
		\draw (\x, \y, 0) -- (\x, \y, 1) ;
		\draw (0.75, \y, \z) -- (5.25, \y, \z) ;
		\draw (\x, 0.75, \z) -- (\x, 2.25, \z) ;
	} } }
	% 生成演算子
	\node[fill, circle] at (2, 2, 1) {} ;
	% 消滅演算子
	\node[fill=white, draw, circle, inner sep=1mm, line width=0.5mm] at (4, 1, 0) {} ;
	% 個数演算子
	\node[fill=white, draw, inner sep=1.2mm, line width=0.5mm] at (3, 1, 1) {} ;
\end{tikzpicture}

\end{minipage}\\
\vspace{1mm}
\begin{minipage}{0.3\linewidth}
\center
$ \sfE _ { 3 } ^ { \p } = $
\begin{tikzpicture}[tdplot_main_coords, scale=0.7, baseline=2.5mm]
	% ボンド	
	\foreach \x in {1, 2, ..., 5} {
	\foreach \y in {1, 2} {
	\foreach \z in {0, 1} {
		\draw (\x, \y, 0) -- (\x, \y, 1) ;
		\draw (0.75, \y, \z) -- (5.25, \y, \z) ;
		\draw (\x, 0.75, \z) -- (\x, 2.25, \z) ;
	} } }
	% 生成演算子
	\node[fill, circle] at (3, 2, 1) {} ;
	% 消滅演算子
	\node[fill=white, draw, circle, inner sep=1mm, line width=0.5mm] at (4, 1, 0) {} ;
	% 個数演算子
	\node[fill=white, draw, inner sep=1.2mm, line width=0.5mm] at (3, 1, 1) {} ;
\end{tikzpicture}

\end{minipage}
\quad
\begin{minipage}{0.3\linewidth}
\center
$ \sfD _ { 3 } ^ { \p } = $
\begin{tikzpicture}[tdplot_main_coords, scale=0.7, baseline=2.5mm]
	% ボンド	
	\foreach \x in {1, 2, ..., 5} {
	\foreach \y in {1, 2} {
	\foreach \z in {0, 1} {
		\draw (\x, \y, 0) -- (\x, \y, 1) ;
		\draw (0.75, \y, \z) -- (5.25, \y, \z) ;
		\draw (\x, 0.75, \z) -- (\x, 2.25, \z) ;
	} } }
	% 生成演算子
	\node[fill, circle] at (3, 2, 1) {} ;
	% 消滅演算子
	\node[fill=white, draw, circle, inner sep=1mm, line width=0.5mm] at (5, 1, 0) {} ;
	% 個数演算子
	\node[fill=white, draw, inner sep=1.2mm, line width=0.5mm] at (3, 1, 1) {} ;
\end{tikzpicture}

\end{minipage}
\vspace{5mm}
\captionsetup{width=0.8\linewidth}
\caption[dummy]{
An example for $ \kb = 3 $ with $ m = 2 $, $ \a = + $, $ \b = - $, $ \s = \up , \t = \down $.
Here, $ \sfD _ { 1 } ^ { \p } $ is generated only from $ \sfA $ and $ \sfE _ { 2 } ^ { \p } $.
$ \sfD _ { 2 } ^ { \p } $ is generated only from $ \sfE _ { 2 } ^ { \p } $ and $ \sfE _ { 3 } ^ { \p } $.
However, $ \sfD _ { 3 } ^ { \p } $ is generated only from $ \sfE _ { 3 } $.
}
\label{F:horizontal condition}
\end{figure}
We find from the commutation relations \eqref{commutation1}, \eqref{commutation2} that
\begin{align}
	\sfD _ { 1 } ^ { \p } = - \b [ \sfA , \n _ { ( \kb , m ) , \up } \n _ { ( \kb , m ) , \down } ] =  [ \sfE _ { 2 } ^ { \p } , \c _ { ( 2 , 1 ) , \s } ^ { \ao } \c _ { ( 1 , 1 ) , \s } ^ { \a } ] ,
\end{align}
which means $ \sfA $ and $ \sfE _ { 2 } ^ { \p } $ generate $ \sfD _ { 1 } ^ { \p } $.
Lemma 3.4 guarantees that these are the only products with possibly nonzero coefficients that generate $ \sfD _ { 1 } ^ { \p } $.
For $ l = 2 , \ldots , \kb - 1 $, we similary have
\begin{align}
	\sfD _ { l } ^ { \p } = [ \sfE _ { l } ^ { \p } , \c _ { ( l + \kb - 2 , m ) , \t } ^ { \bo } \c _ { ( l + \kb - 1 , m ) , \t } ^ { \b } ] = [ \sfE _ { l + 1 } ^ { \p } , \c _ { ( l , 1 ) , \s } ^ { \ao } \c _ { ( l - 1 , 1 ) , \s } ^ { \a } ] ,
\end{align}
and see that $ \sfE _ { l } ^ { \p } $ and $ \sfE _ { l + 1 } ^ { \p } $ are the only relevant products that generate $ \sfD _ { l } ^ { \p } $.
Similarly, we have
\begin{align}
	\sfD _ { \kb } ^ { \p } = [ \sfE _ { \kb } ^ { \p } , \c _ { ( 2 \kb - 2 , m ) , \t } ^ { \bo } \c _ { ( 2 \kb - 1, m ) , \t } ^ { \b } ] ,
\end{align}
and see that $ \sfE _ { \kb } ^ { \prime } $ is the only relevant product that generates $ \sfD _ { \kb } ^ { \p } $.

We then find that the coefficient \eqref{coefficient} for $ \sfD _ { j } ^ { \prime } $ are given by
\begin{align}
	r _ { \sfD _ { 1 } ^ { \p } } &= - \b U q _ { \sfA } + \a t q _ { \sfE _ { 2 } ^ { \prime } } , \label{E:first coefficient in the proof of Lemma 3.5} \\
	r _ { \sfD _ { l } ^ { \p } } &= \b t q _ { \sfE _ { l } ^ { \prime } } + \a t q _ { \sfE _ { l + 1 } ^ { \p } } , \quad l = 2 , \ldots , \kb - 1 , \label{E:second coefficient in the proof of Lemma 3.5} \\
	r _ { \sfD _ { \kb } ^ { \p } } &= \b t q _ { \sfE _ { \kb } ^ { \p } } . \label{E:third coefficient in the proof of Lemma 3.5}
\end{align}
By requiring $ r _ { \sfD _ { j } ^ { \p } } = 0 $ for $ j = 1 , \ldots , \kb $, one readily finds $ q _ { \sfA } = 0 $.
$ \blacksquare $\\

We have thus seen that $ \sfA \in \calPL $ with $ \Wid \sfA = \kb $ may have nonzero $ \qA $ only when it has the form
\begin{align}
	\sfA = \c _ { x , \s } ^ { \a } \c _ { x + ( \kb - 1 ) \ei , \t } ^ { \b } . \label{horizontal operator}
\end{align}
We shall further restrict this.

\begin{lemma}\label{L:same spin}
	For $\kb$ with $ 2 \leq \kb \leq L / 2 $, let $ \sfA $ be of the form \eqref{horizontal operator} with arbitrary $ \a , \b = \pm $, $ \s , \t = \up , \down $, and $ x \in \L $. Then we have $ \qA = 0 $ unless $ \s = \t $.
\end{lemma}
{\em Proof:}\/
We shall show $ \qA = 0 $ assuming $ \s \neq \t $. Without loss of generality, we can set $ \s = \up $, $ \t = \down $. Again, going into the $ d = 2 $ case, and letting $ x = ( 1 , 1 ) $, our $ \sfA $ becomes
\begin{align}
	\sfA = \c _ { ( 1 , 1 ) , \up } ^ { \a } \c _ { ( \kb , 1 ) , \down } ^ { \b } .
\end{align}
To prove $ q _ { \sfA } = 0 $, we further define 
\begin{align}
	\sfD _ { j } ^ { \prime \prime } &= \c _ { ( j , 1 ) , \up } ^ { \a } \n _ { ( \kb , 1 ) , \up } \c _ { ( j + \kb - 1 , 1 ) , \down } ^ { \b } ,
\end{align}
for $ j = 1 , \ldots , \kb - 1 $ and
\begin{align}
	\sfE _ { j } ^ { \prime \prime } &= \c _ { ( j , 1 ) , \up } ^ { \a } \n _ { ( \kb , 1 ) , \up } \c _ { ( j + \kb - 2 , 1 ) , \down } ^ { \b } ,
\end{align}
for $ j = 2 , \ldots , \kb - 1 $. Note that we have $ \Wid \sfA = \Wid \sfD _ { j } ^ { \prime \prime } = \kb $ and $ \Wid \sfE _ { j } ^ { \prime \prime } = \kb - 1 $. See Figure \ref{spin condition}.

\begin{figure}[htb]
\center
\begin{minipage}{0.3\linewidth}
\center
$ \sfA = $
\begin{tikzpicture}[tdplot_main_coords, scale=0.7, baseline=2.5mm]
	% ボンド	
	\foreach \x in {1, 2, ..., 5} {
	\foreach \y in {1, 2} {
	\foreach \z in {0, 1} {
		\draw (\x, \y, 0) -- (\x, \y, 1) ;
		\draw (0.75, \y, \z) -- (5.25, \y, \z) ;
		\draw (\x, 0.75, \z) -- (\x, 2.25, \z) ;
	} } }
	% 生成演算子
	\node[fill, circle] at (1, 2, 1) {} ;
	% 消滅演算子
	\node[fill=white, draw, circle, inner sep=1mm, line width=0.5mm] at (3, 2, 0) {} ;
\end{tikzpicture}

\end{minipage}
\quad
\begin{minipage}{0.3\linewidth}
\center
$ \sfD _ { 1 } ^ { \prime \prime } = $
\begin{tikzpicture}[tdplot_main_coords, scale=0.7, baseline=2.5mm]
	% ボンド	
	\foreach \x in {1, 2, ..., 5} {
	\foreach \y in {1, 2} {
	\foreach \z in {0, 1} {
		\draw (\x, \y, 0) -- (\x, \y, 1) ;
		\draw (0.75, \y, \z) -- (5.25, \y, \z) ;
		\draw (\x, 0.75, \z) -- (\x, 2.25, \z) ;
	} } }
	% 生成演算子
	\node[fill, circle] at (1, 2, 1) {} ;
	% 消滅演算子
	\node[fill=white, draw, circle, inner sep=1mm, line width=0.5mm] at (3, 2, 0) {} ;
	% 個数演算子
	\node[fill=white, draw, inner sep=1.2mm, line width=0.5mm] at (3, 2, 1) {} ;
\end{tikzpicture}

\end{minipage}\\
\vspace{1mm}
\begin{minipage}{0.3\linewidth}
\center
$ \sfE _ { 2 } ^ { \prime \prime } = $
\begin{tikzpicture}[tdplot_main_coords, scale=0.7, baseline=2.5mm]
	% ボンド	
	\foreach \x in {1, 2, ..., 5} {
	\foreach \y in {1, 2} {
	\foreach \z in {0, 1} {
		\draw (\x, \y, 0) -- (\x, \y, 1) ;
		\draw (0.75, \y, \z) -- (5.25, \y, \z) ;
		\draw (\x, 0.75, \z) -- (\x, 2.25, \z) ;
	} } }
	% 生成演算子
	\node[fill, circle] at (2, 2, 1) {} ;
	% 消滅演算子
	\node[fill=white, draw, circle, inner sep=1mm, line width=0.5mm] at (3, 2, 0) {} ;
	% 個数演算子
	\node[fill=white, draw, inner sep=1.2mm, line width=0.5mm] at (3, 2, 1) {} ;
\end{tikzpicture}

\end{minipage}
\quad
\begin{minipage}{0.3\linewidth}
\center
$ \sfD _ { 2 } ^ { \prime \prime } = $
\begin{tikzpicture}[tdplot_main_coords, scale=0.7, baseline=2.5mm]
	% ボンド	
	\foreach \x in {1, 2, ..., 5} {
	\foreach \y in {1, 2} {
	\foreach \z in {0, 1} {
		\draw (\x, \y, 0) -- (\x, \y, 1) ;
		\draw (0.75, \y, \z) -- (5.25, \y, \z) ;
		\draw (\x, 0.75, \z) -- (\x, 2.25, \z) ;
	} } }
	% 生成演算子
	\node[fill, circle] at (2, 2, 1) {} ;
	% 消滅演算子
	\node[fill=white, draw, circle, inner sep=1mm, line width=0.5mm] at (4, 2, 0) {} ;
	% 個数演算子
	\node[fill=white, draw, inner sep=1.2mm, line width=0.5mm] at (3, 2, 1) {} ;
\end{tikzpicture}

\end{minipage}
\vspace{5mm}
\captionsetup{width=0.8\linewidth}
\caption[dummy]{
An example for $ \kb = 3 $ with $ \a = + $, $ \b = - $, and $ \s = \up $. Here, $ \sfD _ { 1 } ^ { \prime \prime } $ is generated only from $ \sfA $ and $ \sfE _ { 2 } ^ { \prime \prime } $, and $ \sfD _ { 2 } ^ { \prime \prime } $ is generated only from $ \sfE _ { 2 } ^ { \prime \prime } $.
}
\label{spin condition}
\end{figure}

Let us be brief since the proof closely resembles that of Lemma 3.5. From the commutation relations \eqref{commutation1}, \eqref{commutation2}, we find
\begin{align}
	\sfD _ { 1 } ^ { \prime \prime } = - \b [ \sfA , \n _ { ( \kb , 1 ) , \up } \n _ { ( \kb , 1 ) , \down } ] = [ \sfE _ { 2 } ^ { \prime \prime } , \c _ { ( 2 , 1 ) , \up } ^ { \ao } \c _ { ( 1 , 1 ) , \up } ^ { \a } ] ,
\end{align}
which means that $ \sfA $ and $ \sfE _ { 2 } ^ { \prime \prime } $ generate $ \sfD _ { 1 } ^ { \prime \prime } $. By Lemma 3.4, these are the only products with possibly nonzero coefficients that generate $ \sfD _ { 1 } ^ { \prime \prime } $. For $ l = 2 , \ldots , \kb - 2 $, we similarly find
\begin{align}
	\sfD _ { l } ^ { \prime \prime } = [ \sfE _ { l } ^ { \prime \prime } , \c _ { ( l + \kb - 2 , 1 ) , \down } ^ { \bo } \c _ { ( l + \kb - 1 , 1 ) , \down } ^ { \b } ] = [ \sfE _ { l + 1 } ^ { \prime \prime } , \c _ { ( l , 1 ) , \up } ^ { \ao } \c _ { ( l - 1 , 1 ) , \up } ^ { \a } ] ,
\end{align}
which shows that $ \sfE _ { l } ^ { \prime \prime } $ and $ \sfE _ { l + 1 } ^ { \prime \prime } $ are the only relevant products that generate $ \sfD _ { l } ^ { \prime \prime } $. Similarly,
\begin{align}
	\sfD _ { \kb - 1 } ^ { \prime \prime } = [ \sfE _ { \kb } ^ { \prime \prime } , \c _ { ( 2 \kb - 3 , 1 ) , \down } ^ { \bo } \c _ { ( 2 \kb - 2 , 1 ) , \down } ^ { \b } ] ,
\end{align}
shows that $ \sfE _ { \kb } ^ { \prime \prime } $ is the only relevant product that generates $ \sfD _ { \kb - 1 } ^ { \prime \prime } $.

From the above, the coefficients \eqref{coefficient} for $ \sfD _ { j } ^ { \prime \prime } $ are found as
\begin{align}
	r _ { \sfD _ { 1 } ^ { \prime \prime } } &= - \b U q _ { \sfA } + \a t q _ { \sfE _ { 2 } ^ { \prime \prime } } , \\
	r _ { \sfD _ { l } ^ { \prime \prime } } &= \b t q _ { \sfE _ { l } ^ { \prime \prime } } + \a t q _ { \sfE _ { l + 1 } ^ { \prime \prime } } , \qquad l = 2 , \ldots , \kb - 2 , \\
	r _ { \sfD _ { \kb - 1 } ^ { \p \p } } &= \b t q _ { \sfE _ { \kb - 1 } ^ { \p \p } } .
\end{align}
By requiring $ r _ { \sfD _ { j } ^ { \prime \prime } } = 0 $ for $ j = 1 , \ldots , \kb - 1 $, we conclude $ q _ { \sfA } = 0 $.
$ \blacksquare $\\

The following lemma finally determines the possible form of $ \sfA \in \calPL $ with $ \Wid \sfA = \kb $ that may have nonzero $ \qA $. The proof is more subtle than the above two.

\begin{lemma}\label{creation-annihilation pair}
	For $\kb$ with $ 2 \leq \kb \leq L / 2 $, let $ \sfA $ be of the form \eqref{horizontal operator} with arbitrary $ \a , \b = \pm $, $ \s = \t = \up , \down $, and $ x \in \L $. Then we have $ \qA = 0 $ unless $ \a \neq \b $.
\end{lemma}

We thus see that $ \sfA $ must be in the particle number preserving form
\begin{align}
	\sfA = \c _ { x , \s } ^ { + } \c _ { x + ( \kb - 1 ) \ei , \s } ^ { - } , \quad \text{ or } \quad \c _ { x , \s } ^ { - } \c _ { x + ( \kb - 1 ) \ei , \s } ^ { + } . \label{standard form0}
\end{align}
Let us call this the {\em standard form} for products with $ \Wid = \kb $.

\medskip
{\em Proof of Lemma \ref{creation-annihilation pair}:}\/
It suffices to treat the case $ \a = \b = + $ since $ \hQ $ is hermitian. We shall show $ q _ { \sfC _ { 1 } ^ { \p \p \p } } = 0 $ for
\begin{align}
	\sfC _ { j } ^ { \p \p \p } = \c _ { ( j , 1 ) , \up } ^ { + } \c _ { ( j + \kb - 1 , 1 ) , \up } ^ { + } ,
\end{align}
with $ j = 1 $ or $ \kb $. We first note that $ \sfC _ { \kb } ^ { \p \p \p } = \calS ^ { \kb - 1 } ( \sfC _ { 1 } ^ { \p \p \p } ) $.
Then \eqref{E:coefficient of Shift} shows
\begin{align}
	q _ { \sfC _ { \kb } ^ { \p \p \p } } = ( - 1 ) ^ { \kb - 1 } q _ { \sfC _ { 1 } ^ { \p \p \p } } . \label{coefficient from Lemma 3.4}
\end{align}
We also define
\begin{align}
	\sfD _ { j } ^ { \p \p \p } &= \c _ { ( j , 1 ) , \up } ^ { + } \n _ { ( \kb , 1 ) , \down } \c _ { ( j + \kb - 1 , 1 ) , \up } ^ { + } ,
\end{align}
for $ j = 1 , \ldots , \kb $, and
\begin{align}
	\sfE _ { j } ^ { \p \p \p } &= \c _ { ( j , 1 ) , \up } ^ { + } \n _ { ( \kb , 1 ) , \down } \c _ { ( j + \kb - 2 , 1 ) , \up } ^ { + } ,
\end{align}
for $ j = 2 , \ldots , \kb $. Note that $ \Wid \sfC _ { j } ^ { \p \p \p } = \Wid \sfD _ { j } ^ { \p \p \p } = \kb $ and $ \Wid \sfE _ { j } ^ { \p \p \p } = \kb - 1 $. See Figures \ref{kb=3, creation-annihilation condition} and \ref{kb=4, creation-annihilation condition}.

\begin{figure}[htb]
\center
\begin{minipage}{0.3\linewidth}
\center
$ \sfC _ { 1 } ^ { \p \p \p } = $
\begin{tikzpicture}[tdplot_main_coords, scale=0.7, baseline=2.5mm]
	% ボンド	
	\foreach \x in {1, 2, ..., 5} {
	\foreach \y in {1, 2} {
	\foreach \z in {0, 1} {
		\draw (\x, \y, 0) -- (\x, \y, 1) ;
		\draw (0.75, \y, \z) -- (5.25, \y, \z) ;
		\draw (\x, 0.75, \z) -- (\x, 2.25, \z) ;
	} } }
	% 生成演算子
	\node[fill, circle] at (1, 2, 1) {} ;
	\node[fill, circle] at (3, 2, 1) {} ;
\end{tikzpicture}

\end{minipage}
\quad
\begin{minipage}{0.3\linewidth}
\center
$ \sfD _ { 1 } ^ { \p \p \p } = $
\begin{tikzpicture}[tdplot_main_coords, scale=0.7, baseline=2.5mm]
	% ボンド	
	\foreach \x in {1, 2, ..., 5} {
	\foreach \y in {1, 2} {
	\foreach \z in {0, 1} {
		\draw (\x, \y, 0) -- (\x, \y, 1) ;
		\draw (0.75, \y, \z) -- (5.25, \y, \z) ;
		\draw (\x, 0.75, \z) -- (\x, 2.25, \z) ;
	} } }
	% 生成演算子
	\node[fill, circle] at (1, 2, 1) {} ;
	\node[fill, circle] at (3, 2, 1) {} ;
	% 個数演算子
	\node[fill=white, draw, inner sep=1.2mm, line width=0.5mm] at (3, 2, 0) {} ;
\end{tikzpicture}

\end{minipage}\\
\vspace{1mm}
\begin{minipage}{0.3\linewidth}
\center
$ \sfE _ { 2 } ^ { \p \p \p } = $
\begin{tikzpicture}[tdplot_main_coords, scale=0.7, baseline=2.5mm]
	% ボンド	
	\foreach \x in {1, 2, ..., 5} {
	\foreach \y in {1, 2} {
	\foreach \z in {0, 1} {
		\draw (\x, \y, 0) -- (\x, \y, 1) ;
		\draw (0.75, \y, \z) -- (5.25, \y, \z) ;
		\draw (\x, 0.75, \z) -- (\x, 2.25, \z) ;
	} } }
	% 生成演算子
	\node[fill, circle] at (2, 2, 1) {} ;
	\node[fill, circle] at (3, 2, 1) {} ;
	% 個数演算子
	\node[fill=white, draw, inner sep=1.2mm, line width=0.5mm] at (3, 2, 0) {} ;
\end{tikzpicture}

\end{minipage}
\quad
\begin{minipage}{0.3\linewidth}
\center
$ \sfD _ { 2 } ^ { \p \p \p } = $
\begin{tikzpicture}[tdplot_main_coords, scale=0.7, baseline=2.5mm]
	% ボンド	
	\foreach \x in {1, 2, ..., 5} {
	\foreach \y in {1, 2} {
	\foreach \z in {0, 1} {
		\draw (\x, \y, 0) -- (\x, \y, 1) ;
		\draw (0.75, \y, \z) -- (5.25, \y, \z) ;
		\draw (\x, 0.75, \z) -- (\x, 2.25, \z) ;
	} } }
	% 生成演算子
	\node[fill, circle] at (2, 2, 1) {} ;
	\node[fill, circle] at (4, 2, 1) {} ;
	% 個数演算子
	\node[fill=white, draw, inner sep=1.2mm, line width=0.5mm] at (3, 2, 0) {} ;
\end{tikzpicture}

\end{minipage}\\
\vspace{1mm}
\begin{minipage}{0.3\linewidth}
\center
$ \sfE _ { 3 } ^ { \p \p \p } = $
\begin{tikzpicture}[tdplot_main_coords, scale=0.7, baseline=2.5mm]
	% ボンド	
	\foreach \x in {1, 2, ..., 5} {
	\foreach \y in {1, 2} {
	\foreach \z in {0, 1} {
		\draw (\x, \y, 0) -- (\x, \y, 1) ;
		\draw (0.75, \y, \z) -- (5.25, \y, \z) ;
		\draw (\x, 0.75, \z) -- (\x, 2.25, \z) ;
	} } }
	% 生成演算子
	\node[fill, circle] at (3, 2, 1) {} ;
	\node[fill, circle] at (4, 2, 1) {} ;
	% 個数演算子
	\node[fill=white, draw, inner sep=1.2mm, line width=0.5mm] at (3, 2, 0) {} ;
\end{tikzpicture}

\end{minipage}
\quad
\begin{minipage}{0.3\linewidth}
\center
$ \sfD _ { 3 } ^ { \p \p \p } = $
\begin{tikzpicture}[tdplot_main_coords, scale=0.7, baseline=2.5mm]
	% ボンド	
	\foreach \x in {1, 2, ..., 5} {
	\foreach \y in {1, 2} {
	\foreach \z in {0, 1} {
		\draw (\x, \y, 0) -- (\x, \y, 1) ;
		\draw (0.75, \y, \z) -- (5.25, \y, \z) ;
		\draw (\x, 0.75, \z) -- (\x, 2.25, \z) ;
	} } }
	% 生成演算子
	\node[fill, circle] at (3, 2, 1) {} ;
	\node[fill, circle] at (5, 2, 1) {} ;
	% 個数演算子
	\node[fill=white, draw, inner sep=1.2mm, line width=0.5mm] at (3, 2, 0) {} ;
\end{tikzpicture}

\end{minipage}\\
\vspace{1mm}
\begin{minipage}{0.3\linewidth}
\center
$ \sfC _ { 3 } ^ { \p \p \p } = $
\begin{tikzpicture}[tdplot_main_coords, scale=0.7, baseline=2.5mm]
	% ボンド	
	\foreach \x in {1, 2, ..., 5} {
	\foreach \y in {1, 2} {
	\foreach \z in {0, 1} {
		\draw (\x, \y, 0) -- (\x, \y, 1) ;
		\draw (0.75, \y, \z) -- (5.25, \y, \z) ;
		\draw (\x, 0.75, \z) -- (\x, 2.25, \z) ;
	} } }
	% 生成演算子
	\node[fill, circle] at (3, 2, 1) {} ;
	\node[fill, circle] at (5, 2, 1) {} ;
\end{tikzpicture}

\end{minipage}
\quad
\begin{minipage}{0.3\linewidth}
\ 
\end{minipage}
\captionsetup{width=0.8\linewidth}
\caption[dummy]{
The products appearing in the proof of Lemma \ref{creation-annihilation pair} for $ \kb = 3 $. 
}
\label{kb=3, creation-annihilation condition}
\end{figure}

\begin{figure}[hbt]
\center
\begin{minipage}{0.4\linewidth}
\center
$ \sfC _ { 1 } ^ { \p \p \p } = $
\begin{tikzpicture}[tdplot_main_coords, scale=0.7, baseline=2.5mm]
	% ボンド	
	\foreach \x in {1, 2, ..., 7} {
	\foreach \y in {1, 2} {
	\foreach \z in {0, 1} {
		\draw (\x, \y, 0) -- (\x, \y, 1) ;
		\draw (0.75, \y, \z) -- (7.25, \y, \z) ;
		\draw (\x, 0.75, \z) -- (\x, 2.25, \z) ;
	} } }
	% 生成演算子
	\node[fill, circle] at (1, 2, 1) {} ;
	\node[fill, circle] at (4, 2, 1) {} ;
\end{tikzpicture}

\end{minipage}
\quad
\begin{minipage}{0.4\linewidth}
\center
$ \sfD _ { 1 } ^ { \p \p \p } = $
\begin{tikzpicture}[tdplot_main_coords, scale=0.7, baseline=2.5mm]
	% ボンド	
	\foreach \x in {1, 2, ..., 7} {
	\foreach \y in {1, 2} {
	\foreach \z in {0, 1} {
		\draw (\x, \y, 0) -- (\x, \y, 1) ;
		\draw (0.75, \y, \z) -- (7.25, \y, \z) ;
		\draw (\x, 0.75, \z) -- (\x, 2.25, \z) ;
	} } }
	% 生成演算子
	\node[fill, circle] at (1, 2, 1) {} ;
	\node[fill, circle] at (4, 2, 1) {} ;
	% 個数演算子
	\node[fill=white, draw, inner sep=1.2mm, line width=0.5mm] at (4, 2, 0) {} ;
\end{tikzpicture}

\end{minipage}\\
\vspace{-0.5mm}
\begin{minipage}{0.4\linewidth}
\center
$ \sfE _ { 2 } ^ { \p \p \p } = $
\begin{tikzpicture}[tdplot_main_coords, scale=0.7, baseline=2.5mm]
	% ボンド	
	\foreach \x in {1, 2, ..., 7} {
	\foreach \y in {1, 2} {
	\foreach \z in {0, 1} {
		\draw (\x, \y, 0) -- (\x, \y, 1) ;
		\draw (0.75, \y, \z) -- (7.25, \y, \z) ;
		\draw (\x, 0.75, \z) -- (\x, 2.25, \z) ;
	} } }
	% 生成演算子
	\node[fill, circle] at (2, 2, 1) {} ;
	\node[fill, circle] at (4, 2, 1) {} ;
	% 個数演算子
	\node[fill=white, draw, inner sep=1.2mm, line width=0.5mm] at (4, 2, 0) {} ;
\end{tikzpicture}

\end{minipage}
\quad
\begin{minipage}{0.4\linewidth}
\center
$ \sfD _ { 2 } ^ { \p \p \p } = $
\begin{tikzpicture}[tdplot_main_coords, scale=0.7, baseline=2.5mm]
	% ボンド	
	\foreach \x in {1, 2, ..., 7} {
	\foreach \y in {1, 2} {
	\foreach \z in {0, 1} {
		\draw (\x, \y, 0) -- (\x, \y, 1) ;
		\draw (0.75, \y, \z) -- (7.25, \y, \z) ;
		\draw (\x, 0.75, \z) -- (\x, 2.25, \z) ;
	} } }
	% 生成演算子
	\node[fill, circle] at (2, 2, 1) {} ;
	\node[fill, circle] at (5, 2, 1) {} ;
	% 個数演算子
	\node[fill=white, draw, inner sep=1.2mm, line width=0.5mm] at (4, 2, 0) {} ;
\end{tikzpicture}

\end{minipage}\\
\vspace{-0.5mm}
\begin{minipage}{0.4\linewidth}
\center
$ \sfE _ { 3 } ^ { \p \p \p } = $
\begin{tikzpicture}[tdplot_main_coords, scale=0.7, baseline=2.5mm]
	% ボンド	
	\foreach \x in {1, 2, ..., 7} {
	\foreach \y in {1, 2} {
	\foreach \z in {0, 1} {
		\draw (\x, \y, 0) -- (\x, \y, 1) ;
		\draw (0.75, \y, \z) -- (7.25, \y, \z) ;
		\draw (\x, 0.75, \z) -- (\x, 2.25, \z) ;
	} } }
	% 生成演算子
	\node[fill, circle] at (3, 2, 1) {} ;
	\node[fill, circle] at (5, 2, 1) {} ;
	% 個数演算子
	\node[fill=white, draw, inner sep=1.2mm, line width=0.5mm] at (4, 2, 0) {} ;
\end{tikzpicture}

\end{minipage}
\quad
\begin{minipage}{0.4\linewidth}
\center
$ \sfD _ { 3 } ^ { \p \p \p } = $
\begin{tikzpicture}[tdplot_main_coords, scale=0.7, baseline=2.5mm]
	% ボンド	
	\foreach \x in {1, 2, ..., 7} {
	\foreach \y in {1, 2} {
	\foreach \z in {0, 1} {
		\draw (\x, \y, 0) -- (\x, \y, 1) ;
		\draw (0.75, \y, \z) -- (7.25, \y, \z) ;
		\draw (\x, 0.75, \z) -- (\x, 2.25, \z) ;
	} } }
	% 生成演算子
	\node[fill, circle] at (3, 2, 1) {} ;
	\node[fill, circle] at (6, 2, 1) {} ;
	% 個数演算子
	\node[fill=white, draw, inner sep=1.2mm, line width=0.5mm] at (4, 2, 0) {} ;
\end{tikzpicture}

\end{minipage}\\
\vspace{-0.5mm}
\begin{minipage}{0.4\linewidth}
\center
$ \sfE _ { 4 } ^ { \p \p \p } = $
\begin{tikzpicture}[tdplot_main_coords, scale=0.7, baseline=2.5mm]
	% ボンド	
	\foreach \x in {1, 2, ..., 7} {
	\foreach \y in {1, 2} {
	\foreach \z in {0, 1} {
		\draw (\x, \y, 0) -- (\x, \y, 1) ;
		\draw (0.75, \y, \z) -- (7.25, \y, \z) ;
		\draw (\x, 0.75, \z) -- (\x, 2.25, \z) ;
	} } }
	% 生成演算子
	\node[fill, circle] at (4, 2, 1) {} ;
	\node[fill, circle] at (6, 2, 1) {} ;
	% 個数演算子
	\node[fill=white, draw, inner sep=1.2mm, line width=0.5mm] at (4, 2, 0) {} ;
\end{tikzpicture}

\end{minipage}
\quad
\begin{minipage}{0.4\linewidth}
\center
$ \sfD _ { 4 } ^ { \p \p \p } = $
\begin{tikzpicture}[tdplot_main_coords, scale=0.7, baseline=2.5mm]
	% ボンド	
	\foreach \x in {1, 2, ..., 7} {
	\foreach \y in {1, 2} {
	\foreach \z in {0, 1} {
		\draw (\x, \y, 0) -- (\x, \y, 1) ;
		\draw (0.75, \y, \z) -- (7.25, \y, \z) ;
		\draw (\x, 0.75, \z) -- (\x, 2.25, \z) ;
	} } }
	% 生成演算子
	\node[fill, circle] at (4, 2, 1) {} ;
	\node[fill, circle] at (7, 2, 1) {} ;
	% 個数演算子
	\node[fill=white, draw, inner sep=1.2mm, line width=0.5mm] at (4, 2, 0) {} ;
\end{tikzpicture}

\end{minipage}\\
\vspace{-0.5mm}
\begin{minipage}{0.4\linewidth}
\center
$ \sfC _ { 4 } ^ { \p \p \p } = $
\begin{tikzpicture}[tdplot_main_coords, scale=0.7, baseline=2.5mm]
	% ボンド	
	\foreach \x in {1, 2, ..., 7} {
	\foreach \y in {1, 2} {
	\foreach \z in {0, 1} {
		\draw (\x, \y, 0) -- (\x, \y, 1) ;
		\draw (0.75, \y, \z) -- (7.25, \y, \z) ;
		\draw (\x, 0.75, \z) -- (\x, 2.25, \z) ;
	} } }
	% 生成演算子
	\node[fill, circle] at (4, 2, 1) {} ;
	\node[fill, circle] at (7, 2, 1) {} ;
\end{tikzpicture}

\end{minipage}
\quad
\begin{minipage}{0.4\linewidth}
\ 
\end{minipage}
\captionsetup{width=0.8\linewidth}
\caption[dummy]{
The products appearing in the proof of Lemma \ref{creation-annihilation pair} for $ \kb = 4 $.
}
\label{kb=4, creation-annihilation condition}
\end{figure}

\noindent In the same manner as in the proofs of Lemmas 3.5 and 3.6, the commutation relations \eqref{commutation1} and \eqref{commutation2} yield
\eqa
\sfD _ { 1 } ^ { \p \p \p } &= - [ \sfC _ { 1 } ^ { \p \p \p } , \n _ { ( \kb , 1 ) , \up } \n _ { ( \kb , 1 ) , \down } ] = - [ \sfE _ { 2 } ^ { \p \p \p } , \c _ { ( 1 , 1 ) , \up } ^ { + } \c _ { ( 2 , 1 ) , \up } ^ { - } ] , \\
\sfD _ { l } ^ { \p \p \p } &= - [ \sfE _ { l } ^ { \p \p \p } , \c _ { ( l + \kb - 1 , 1 ) , \up } ^ { + } \c _ { ( l + \kb - 2 , 1 ) , \up } ^ { - } ] = - [ \sfE _ { l + 1 } ^ { \p \p \p } , \c _ { ( l , 1 ) , \up } ^ { + } \c _ { ( l + 1 , 1 ) ,\up } ^ { - } ] , \qquad l = 2 , \ldots , \kb - 1 , \\
\sfD _ { \kb } ^ { \p \p \p } &= - [ \sfE _ { \kb } ^ { \p \p \p } , \c _ { ( \kb - 1 , 1 ) , \up } ^ { + } \c _ { ( \kb , 1 ) , \up } ^ { - } ] = - [ \sfC _ { \kb } ^ { \p \p \p } , \n _ { ( \kb , 1 ) , \up } \n _ { ( \kb , 1 ) , \down } ] ,
\ena
from which we see that $ \sfC _ { 1 } ^ { \p \p \p } $ and $ \sfE _ { 2 } ^ { \p \p \p } $ are the only relevant products that generate $ \sfD _ { 1 } ^ { \p \p \p } $. Likewise, $ \sfE _ { l } ^ { \p \p \p } $ and $ \sfE _ { l + 1 } ^ { \p \p \p } $ are the only relevant products that generate $ \sfD _ { l } ^ { \p \p \p } $ for $ l = 2 , \ldots , \kb - 1 $, and $ \sfE _ { \kb } ^ { \p \p \p } $ and $ \sfC _ { \kb } ^ { \p \p \p } $ are the only relevant products that generate $ \sfD _ { \kb } ^ { \p \p \p } $.

We then find that the coefficients \eqref{coefficient} for $ \sfD _ { j } ^ { \p \p \p } $ are given by
\begin{align}
r _ { \sfD _ { 1 } ^ { \p \p \p } } &= - U q _ { \sfC _ { 1 } ^ { \p \p \p } } + t q _ { \sfE _ { 2 } ^ { \p \p \p } } , \label{E:first coefficient in the proof of Lemma 3.7} \\
r _ { \sfD _ { l } ^ { \p \p \p } } &= t q _ { \sfE _ { l } ^ { \p \p \p } } + t q _ { \sfE _ { l + 1 } ^ { \p \p \p } } , \qquad l = 2 , \ldots , \kb - 1 , \label{E:second coefficient in the proof of Lemma 3.7} \\
r _ { \sfD _ { \kb } ^ { \p \p \p } } &= t q _ { \sfE _ { \kb } ^ { \p \p \p } } - U q _ { \sfC _ { \kb } ^ { \p \p \p } } . \label{E:third coefficient in the proof of Lemma 3.7}
\end{align}
By requiring $ r _ { \sfD _ { j } ^ { \p \p \p } } = 0 $ for $ j = 1 , \ldots , \kb $, we obtain
\begin{align}
	q _ { \sfC _ { \kb } ^ { \p \p \p } } = ( - 1 ) ^ { \kb } q _ { \sfC _ { 1 } ^ { \p \p \p } } .
\end{align}
Comparing this result with \eqref{coefficient from Lemma 3.4}, we find $ q _ { \sfC _ { 1 } ^ { \p \p \p } } = 0 $.
$ \blacksquare $\\

Recall that, to prove Theorem \ref{main theorem}, it is sufficient to show $q_{\sfA}=0$ for any $\sfA\in\calPL$ with $\Wid\sfA=\kb$.
Lemmas~\ref{L:two site product}--\ref{creation-annihilation pair} show that we only need to prove $ \qA = 0 $ for $ \sfA $ of the standard form \eqref{standard form0} with $ 3 \leq \kb \leq L / 2 $.
Unlike in the case of spin systems treated in \cite{ShiraishiTasaki2024, FutamiTasaki2025}, however, this task requires us to further characterize products with width $ \kb - 1 $ that may contribute to a nontrivial local conserved quantity.\footnote{%
A similar case is found in Section 4.4.3 of \cite{Shiraishi2025}, where a quantum spin chain that shares some features with the Hubbard model is treated.
\label{fn:Shiraishi}
}
One may attempt to obtain further relations among the coefficients of $ \sfA $ with $ \Wid \sfA = \kb $ by taking the commutator with the hopping Hamiltonian, which generates $ \sfB $ with $ \Wid \sfB = \kb $.
However, such relations turn out not to be helpful for the present proof.
The following lemma provides necessary characterizations.

\begin{lemma}\label{substandard form}
	For $\kb$ with $ 3 \leq \kb \leq L / 2 $, let $ \sfA \in \calPL $ be such that $ \Wid \sfA = \kb - 1 $. Then we have $ \qA = 0 $ unless $ \sfA $ is
\eq
\sfA = \c _ { x , \s } ^ { \a } \c _ { y , \t } ^ { \b } , \label{two site product2}
\en
with some $ \a , \b = \pm $, $ \s , \t = \up , \down $ and $ x , y \in \L $ such that $ \big ( y - x \big ) _ { 1 } = \kb - 2 $,
\eq
\sfA = \c _ { x , \s } ^ { \a } \n _ { x + m \ei , \so } \c _ { x + ( \kb - 2 ) \ei , \s } ^ { \b } , \label{substandard form1}
\en
with some $ \a , \b = \pm $ such that $ \a \b = - 1 $, $ \s = \up , \down $, $ x \in \L $, and $ m = 0 , \ldots , \kb - 2 $, or,
\eq
\sfA = \c _ { x , \s } ^ { \a } \c _ { x , + m \ei , \s } ^ { \ao } \c _ { x + m \ei , \so } ^ { \bo } \c _ { x + ( \kb - 2 ) \ei , \so } ^ { \b } , \label{step form}
\en
with some $ \a , \b = \pm $, $ \s = \up , \down $, $ x \in \Lambda $, and $ m = 1 , \ldots , \kb - 3 $.
\end{lemma}

With an extra effort, we can further restrict \eqref{two site product2} to products with $ \a \b = - 1 $, $ \s = \t $, and $ y = x + ( \kb - 2 ) \ei $.
But the present lemma is sufficient for us.

It is worth pointing out that the standard form \eqref{standard form0} for the width $\kb$ products and the above \eqref{two site product2}, \eqref{substandard form1} (with the further restriction mentioned above) for the width $\kb-1$ products precisely recover the leading terms of the conserved quantity (with the maximum width $\kb$) of the one-dimensional Hubbard model.
See, e.g., Theorems~1 and 2 of \cite{Grosse1989}.
It is interesting that the (near) precise form of the exact conserved quantity for the one-dimensional Hubbard model is necessary for the proof of the absence of conserved quantities in higher dimensions.

\medskip
\noindent{\em Proof:}\/
Let us emphasize that the proof of the present lemma is essentially different from that of (similarly looking) Lemma \ref{leftright-most site conditions} since, in the present case, there are products with possibly nonzero coefficients whose width is strictly larger than that of the product in consideration.

Consider a product $ \sfA \in \calPL $ with $ \Wid \sfA = \kb - 1 $.
Let $ \xr \in \Supp \sfA $ be the right-most site of $ \sfA $.
As in \eqref{definition of commutator manner} in the proof of Lemma 3.2, define a product $ \sfB \in \calPL $ as a nonzero product written as
\begin{align}
	\sfB = \pm [ \sfA , \cp _ { \xr , \s } \cm _ { \xr + \ei , \s } ] \quad \text{or} \quad \sfB = \pm [ \sfA , \cp _ { \xr + \ei , \s } \cm _ { \xr , \s } ] . \label{aaa}
\end{align}
Since the new site $ \xr + \ei $ is added to $ \Supp \sfA $, we have $ \Wid \sfB = \kb $.
We then ask if there exists another product $ \sfA ^ { \prime } $ with $ \Wid \sfA ^ { \prime } \leq \kb $ that generates $ \sfB $.
This is the point where the situation differs from that in Lemma \ref{leftright-most site conditions}.

We shall consider the following two cases:

\noindent case 1. \ There exists a product with width $ \kb $ that generates $ \sfB $.

\noindent case 2. \ There does not exist a product with width $ \kb $ that generates $ \sfB $.

In case 1, by Lemmas \ref{leftright-most site conditions}--\ref{creation-annihilation pair}, products with width $ \kb $ whose coefficients may be nonzero are limited to the standard form \eqref{standard form0}.
Thus, if $ \sfB $ is generated by a commutator with the hopping term $ \Hhop $, then
\eq
\sfB = \c _ { \xr - ( \kb - 2 ) \ei \pm \bm{ e } _ { n } , \s } ^ { \a } \c _ { \xr + \ei , \s } ^ { \b } , \quad \text{or} \quad \sfB = \c _ { \xr - ( \kb - 2 ) \ei , \s } ^ { \a } \c _ { \xr + \ei \pm \bm{ e } _ { n } , \s } ^ { \b } , \label{standard form with hopping}
\en
with some $ \a , \b = \pm $ such that $ \a \b = - 1 $, $ \s = \up , \down $ and $ n = 2 , \ldots , d $.
Here, the $ \pm $ signs are not taken consistently.
If $ \sfB $ is generated by a commutator with the interaction term $ \Hint $, then
\eq
\sfB = \c _ { \xr - ( \kb - 2 ) \ei , \s } ^ { \a } \n _ { \xr - ( \kb - 2 ) \ei , \so } \c _ { \xr + \ei , \s } ^ { \b } , \label{standard form with interaction}
\en
with some $ \a , \b = \pm $ such that $ \a \b = - 1 $ and $ n = 2 , \ldots , d $.
Note that, unlike in \eqref{standard form with interaction}, we only need to take the commutator at the left-most site since we already know that $ \sfB $ has $ \hB _ { \xr + \ei } = \c _ { \xr + \ei , \s } ^ { \b } $ at its right-most site $ \xr + \ei $.

If $ \sfB $ is of the form \eqref{standard form with hopping}, then by examining commutation relations, we find that $ \sfA $ must be
\eq
\sfA = \c _ { \xr - ( \kb - 2 ) \ei \pm \bm{ e } _ { n } , \s } ^ { \a } \c _ { \xr , \s } ^ { \b } , \label{two site product3}
\en
with some $ \a , \b = \pm $ such that $ \a \b = - 1 $ and $ \s = \up , \down $.
This corresponds to \eqref{two site product2} with $ x = \xr - ( \kb - 2 ) \ei \pm \bm{ e } _ { n } $, $ y = \xr $, and $ \s = \t = \up , \down $.
If $ \sfB $ is of the form \eqref{standard form with interaction}, on the otherhand, then the commutation relations \eqref{aaa} imply that
\eq
\sfA = \c _ { \xr - ( \kb - 2 ) \ei , \s } ^ { \a } \n _ { \xr - ( \kb - 2 ) \ei , \so } \c _ { \xr , \s } ^ { \b } , \label{substandard form2}
\en
with some $ \a , \b = \pm $ such that $ \a \b = - 1 $ and $ \s = \up , \down $.
This corresponds to \eqref{substandard form1} with $ m = 0 $.
This completes the proof for case 1.

For case 2, we can exactly repeat the proof of Lemma 3.2 (with $ \kb $ replaced with $ \kb - 1 $) to see that $ \qA = 0 $ unless $ \sfA $ satisfies (i) of Lemma 3.2.
We can further proceed as before (still $ \kb $ replaced with $ \kb - 1 $) to see that $ \qA = 0 $ unless the Shiraishi shift $ \calS ( \sfA ) $ of $ \sfA $ exists.
If $ \calS ( \sfA ) $ exists, we have $ q _ { \calS ( \sfA ) } = \qA $.
We of course have $ \Wid \calS ( \sfA ) = \kb - 1 $.

We then return to the beginning of the proof with $ \sfA $ replaced by $ \calS ( \sfA ) $.
If this falls into case 1, then we see that $ \calS ( \sfA ) $ is of the form \eqref{two site product2} or \eqref{substandard form1} with $ m = 0 $.
This shows $ \sfA $ is of the form \eqref{two site product2},  \eqref{substandard form1} with $ m = 1 $, or \eqref{step form} with $ m = 1 $.\footnote{An inspection shows that the form \eqref{two site product2} is indeed impossible.\label{a}}
For case 2, we again see that $ q _ { \calS ( \sfA ) } = 0 $ (and hence $ \qA = 0 $) unless the second Shiraishi shift $ \calS ^ { 2 } ( \sfA ) $ exists.

Clearly this process can be repeated.
If $ \sfA , \calS ( \sfA ) , \ldots , \calS ^ { m - 1 } ( \sfA ) $ fall into case 2 and $ \calS ^ { m } ( \sfA ) $ falls into case 1 for the first time with $ m < \kb $, one finds $ \sfA $ is of the form \eqref{substandard form1} or \eqref{step form}.\footnote{See footnote~\ref{a}.}
If it happens that all of $ \sfA , \calS ( \sfA ) , \ldots , \calS ^ { \kb - 1 } ( \sfA ) $ fall into case 2, then one finds from the same argument as in the proof of Lemma 3.4 that $ \sfA $ has the form \eqref{two site product2}.
$ \blacksquare $\\

We are now ready for the proof of our main result, Theorem 2.1.
As we noted above Lemma 3.8, our goal is to show $ \qA = 0 $ for all $ \sfA $ of the standard form \eqref{standard form0} with $ 3 \leq \Wid \sfA \leq L / 2 $.

Again, we can reduce this, without losing generality, to showing $ q _ { \hC _ { 1 } } = 0 $ where
\begin{align}
	\hC _ { 1 } = \c _ { ( 1 , 1 ) , \up } ^ { + } \c _ { ( \kb , 1 ) , \up } ^ { - } . \label{standard form}
\end{align}
By taking the commutator between $ \sfC _ { 1 } $ and the interaction term at the right-most site of $ \sfC _ { 1 } $, we get
\begin{align}
	\sfD _ { 1 } = [ \sfC _ { 1 } , \n _ { ( \kb , 1 ) ,\up } \n _ { ( \kb , 1 ) , \down } ] = \cp _ { ( 1 , 1 ) , \up } \cm _ { ( \kb , 1 ) , \up } \n _ { ( \kb , 1 ) , \down } ,
\end{align}
which has $ \Wid \sfD _ { 1 } = \kb $.
It is clear that $ \sfC _ { 1 } $ is the only product with possibly nonzero coefficient with $ \Wid = \kb $ that generates $ \sfD _ { 1 } $.
Let us then define
\begin{align}
	\sfE _ { 2 } = \cp _ { ( 2 , 1 ) , \up } \cm _ { ( \kb , 1 ) , \up } \n _ { ( \kb , 1 ) , \down } ,
\end{align}
which satisfies $ \Wid \sfE _ { 2 } = \kb - 1 $ and
\begin{align}
	\sfD _ { 1 } = - [ \sfE _ { 2 } , \cp _ { ( 1 , 1 ) , \up } \cm _ { ( 2 , 1 ) , \up } ] .
\end{align}
Lemma 3.8 guarantees that $ \sfE _ { 2 } $ is the only relevant product with $ \Wid = \kb - 1 $ that generates $ \sfD _ { 1 } $.
We thus find that the coefficient \eqref{coefficient} for $ \sfD _ { 1 } $ is
\begin{align}
	r _ { \sfD _ { 1 } } = U q _ { \sfC _ { 1 } } + t q _ { \sfE _ { 2 } } .
\end{align}
By requiring $ r _ { \sfD _ { 1 } } = 0 $, we get
\begin{align}
	q _ { \sfE _ { 2 } } = - \frac{ U }{ t } q _ { \sfC _ { 1 } } .
\end{align}
We thus see it suffices to show $ q _ { \sfE _ { 2 } } = 0 $ to prove the desired $ q _ { \sfC _ { 1 } } = 0 $.

Let us generalize the consideration and define
\begin{align}
	\sfD _ { j } &= \cp _ { ( j , 1 ) , \up } \n _ { ( \kb , 1 ) , \down } \cm _ { ( j + \kb - 1 , 1 ) , \up } ,
\end{align}
for $ j = 1 , \ldots , \kb $, and
\begin{align}
	\sfE _ { j } = \cp _ { ( j , 1 ) , \up } \n _ { ( \kb - 1 , 1 ) , \down } \cm _ { ( j + \kb - 2 , 1 ) , \up } ,
\end{align}
for $ j = 2 , \ldots , \kb $. See Figure \ref{coefficient of substandard form7}. Note that $ \Wid \sfD _ { j } = \kb $ and $ \Wid \sfE _ { j } = \kb - 1 $. Then it is verified that
\begin{align}
	\sfD _ { l } = [ \sfE _ { l } , \cp _ { ( l + \kb - 2 , 1 ) , \up } \cm _ { ( l + \kb - 1 , 1 ) , \up } ] = - [ \sfE _ { l + 1 } , \cp _ { ( l - 1 , 1 ) , \up } \cm _ { ( l , 1 ) , \up } ] ,
\end{align}
for $ l = 2 , \ldots , \kb - 1 $.
We see that $ \sfE _ { l } $ and $ \sfE _ { l + 1 } $ are the only products with possibly nonzero coefficients that generate $ \sfD _ { l } $, and hence the coefficient \eqref{coefficient} for $ \sfD _ { l } $ is
\begin{align}
	r _ { \sfD _ { l } } = - t q _ { \sfE _ { l } } + t q _ { \sfE _ { l + 1 } } ,
\end{align}
for $ l = 2 , \ldots , \kb - 1 $. By requiring $ r _ { \sfD _ { l } } = 0 $, we see that $ q _ { \sfE _ { j } } $ is independent of $ j = 2 , \ldots , \kb $.

Let us summarize the observation as the following lemma.\\

\begin{figure}[htb]
\center
\begin{minipage}{0.3\linewidth}
\center
$ \sfC _ { 1 } = $
\begin{tikzpicture}[tdplot_main_coords, scale=0.7, baseline=2.5mm]
	% ボンド	
	\foreach \x in {1, 2, ..., 5} {
	\foreach \y in {1, 2} {
	\foreach \z in {0, 1} {
		\draw (\x, \y, 0) -- (\x, \y, 1) ;
		\draw (0.75, \y, \z) -- (5.25, \y, \z) ;
		\draw (\x, 0.75, \z) -- (\x, 2.25, \z) ;
	} } }
	% 生成演算子
	\node[fill, circle] at (1, 2, 1) {} ;
	% 消滅演算子
	\node[fill=white, draw, circle, inner sep=1mm, line width=0.5mm] at (3, 2, 1) {} ;
\end{tikzpicture}

\end{minipage}
\quad
\begin{minipage}{0.3\linewidth}
\center
$ \sfD _ { 1 } = $
\begin{tikzpicture}[tdplot_main_coords, scale=0.7, baseline=2.5mm]
	% ボンド	
	\foreach \x in {1, 2, ..., 5} {
	\foreach \y in {1, 2} {
	\foreach \z in {0, 1} {
		\draw (\x, \y, 0) -- (\x, \y, 1) ;
		\draw (0.75, \y, \z) -- (5.25, \y, \z) ;
		\draw (\x, 0.75, \z) -- (\x, 2.25, \z) ;
	} } }
	% 生成演算子
	\node[fill, circle] at (1, 2, 1) {} ;
	% 消滅演算子
	\node[fill=white, draw, circle, inner sep=1mm, line width=0.5mm] at (3, 2, 1) {} ;
	% 個数演算子
	\node[fill=white, draw, inner sep=1.2mm, line width=0.5mm] at (3, 2, 0) {} ;
\end{tikzpicture}

\end{minipage}\\
\vspace{1mm}
\begin{minipage}{0.3\linewidth}
\center
$ \sfE _ { 2 } = $
\begin{tikzpicture}[tdplot_main_coords, scale=0.7, baseline=2.5mm]
	% ボンド	
	\foreach \x in {1, 2, ..., 5} {
	\foreach \y in {1, 2} {
	\foreach \z in {0, 1} {
		\draw (\x, \y, 0) -- (\x, \y, 1) ;
		\draw (0.75, \y, \z) -- (5.25, \y, \z) ;
		\draw (\x, 0.75, \z) -- (\x, 2.25, \z) ;
	} } }
	% 生成演算子
	\node[fill, circle] at (2, 2, 1) {} ;
	% 消滅演算子
	\node[fill=white, draw, circle, inner sep=1mm, line width=0.5mm] at (3, 2, 1) {} ;
	% 個数演算子
	\node[fill=white, draw, inner sep=1.2mm, line width=0.5mm] at (3, 2, 0) {} ;
\end{tikzpicture}

\end{minipage}
\quad
\begin{minipage}{0.3\linewidth}
\center
$ \sfD _ { 2 } = $
\begin{tikzpicture}[tdplot_main_coords, scale=0.7, baseline=2.5mm]
	% ボンド	
	\foreach \x in {1, 2, ..., 5} {
	\foreach \y in {1, 2} {
	\foreach \z in {0, 1} {
		\draw (\x, \y, 0) -- (\x, \y, 1) ;
		\draw (0.75, \y, \z) -- (5.25, \y, \z) ;
		\draw (\x, 0.75, \z) -- (\x, 2.25, \z) ;
	} } }
	% 生成演算子
	\node[fill, circle] at (2, 2, 1) {} ;
	% 消滅演算子
	\node[fill=white, draw, circle, inner sep=1mm, line width=0.5mm] at (4, 2, 1) {} ;
	% 個数演算子
	\node[fill=white, draw, inner sep=1.2mm, line width=0.5mm] at (3, 2, 0) {} ;
\end{tikzpicture}

\end{minipage}\\
\vspace{1mm}
\begin{minipage}{0.3\linewidth}
\center
$ \sfE _ { 3 } = $
\begin{tikzpicture}[tdplot_main_coords, scale=0.7, baseline=2.5mm]
	% ボンド	
	\foreach \x in {1, 2, ..., 5} {
	\foreach \y in {1, 2} {
	\foreach \z in {0, 1} {
		\draw (\x, \y, 0) -- (\x, \y, 1) ;
		\draw (0.75, \y, \z) -- (5.25, \y, \z) ;
		\draw (\x, 0.75, \z) -- (\x, 2.25, \z) ;
	} } }
	% 生成演算子
	\node[fill, circle] at (3, 2, 1) {} ;
	% 消滅演算子
	\node[fill=white, draw, circle, inner sep=1mm, line width=0.5mm] at (4, 2, 1) {} ;
	% 個数演算子
	\node[fill=white, draw, inner sep=1.2mm, line width=0.5mm] at (3, 2, 0) {} ;
\end{tikzpicture}

\end{minipage}
\quad
\begin{minipage}{0.3\linewidth}
\ 
\end{minipage}
\vspace{3mm}
\captionsetup{width=0.8\linewidth}
\caption[dummy]{
The products $ \sfC _ { 1 } $, $ \sfD _ { 1 } $ and $ \sfE _ { j } $ for $ \kb = 3 $.
Here, $ \sfD _ { 1 } $ is generated only by $ \sfC _ { 1 } $ and $ \sfE _ { 2 } $, $ \sfD _ { 2 } $ is generated only by $ \sfE _ { 2 } $ and $ \sfE _ { 3 } $.
}
\label{coefficient of substandard form7}
\end{figure}

\begin{lemma}\label{coefficient of substandard form}
	For $\kb$ with $ 3 \leq \kb \leq L / 2 $, let 
	\begin{align}
		\sfC _ { 1 } &= \cp _ { ( 1 , 1 ) , \up } \cm _ { ( \kb , 1 ) , \up } , \\
		\sfE _ { j } &= \cp _ { ( j , 1 ) , \up } \n _ { ( \kb , 1 ) , \down } \cm _ { ( j + \kb - 2 , 1 ) , \up } ,
	\end{align}
	for $ j = 2 , \ldots , \kb $. We then have for any $ j = 2 , \ldots , \kb $ that
	\begin{align}
		q _ { \sfE _ { j } } = - \frac{ U }{ t } q _ { \sfC _ { 1 } } .
	\end{align}
\end{lemma}

\subsection{Third step: basic relations for products with width $ \kb - 1 $}

We are ready to complete our proof.
Here we shall make use of relations that generate products with $ \Wid = \kb - 1 $ to show that $ q _ { \sfE _ { j } } = 0 $.
This implies the desired $ q _ { \sfC _ { 1 } } = 0 $ and hence the main theorem, Theorem 2.1.

We shall treat general $ \kb $ with $ 3 \leq \kb \leq L / 2 $.
Since the case with $ \kb = 3 $ is exceptional, we shall treat the cases with $ \kb = 3 $ and 4, before writing down the proof for general $ \kb \geq 4 $.

\subsubsection{The case with $ \kb = 3 $}

Let us define
\eqa
\sfE _ { 2 } &= \cp _ { ( 2 , 1 ) , \up } \n _ { ( 3 , 1 ) , \down } \cm _ { ( 3 , 1 ) , \up } , & \sfE _ { 3 } &= \cp _ { ( 3 , 1 ) , \up } \n _ { ( 3 , 1 ) , \down } \cm _ { ( 4 , 1 ) , \up } , \\
\sfF _ { 2 } &= \cp _ { ( 2 , 1 ) , \up } \cm _ { ( 3 , 1 ) , \up } \cp _ { ( 3 , 1 ) , \down } \cm _ { ( 3 , 2 ) , \down } , & \sfF _ { 3 } &= \cp _ { ( 3 , 1 ) , \up } \cp _ { ( 3 , 1 ) , \down } \cm _ { ( 3 , 2 ) , \down } \cm _ { ( 4 , 1 ) , \up } , \\
& & \sfG _ { 3 } &= \n _ { ( 3 , 1 ) , \up } \cp _ { ( 3 , 1 ) , \down } \cm _ { ( 3 , 2 ) , \down } ,
\ena
where $ \Wid \sfE _ { j } = \Wid \sfF _ { j } = 2 = \kb - 1 $ and $ \Wid \sfG _ { j } = 1 = \kb - 2 $. See Figure \ref{high dimensional effect case k = 3}.

\begin{figure}[htb]
\center
\begin{minipage}{0.3\linewidth}
\center
$ \sfC _ { 1 } = $
\begin{tikzpicture}[tdplot_main_coords, scale=0.7, baseline=2.5mm]
	% ボンド	
	\foreach \x in {1, 2, ..., 5} {
	\foreach \y in {1, 2} {
	\foreach \z in {0, 1} {
		\draw (\x, \y, 0) -- (\x, \y, 1) ;
		\draw (0.75, \y, \z) -- (5.25, \y, \z) ;
		\draw (\x, 0.75, \z) -- (\x, 2.25, \z) ;
	} } }
	% 生成演算子
	\node[fill, circle] at (1, 2, 1) {} ;
	% 消滅演算子
	\node[fill=white, draw, circle, inner sep=1mm, line width=0.5mm] at (3, 2, 1) {} ;
\end{tikzpicture}

\end{minipage}
\quad
\begin{minipage}{0.3\linewidth}
\ 
\end{minipage}\\
\vspace{1mm}
\begin{minipage}{0.3\linewidth}
\center
$ \sfE _ { 2 } = $
\begin{tikzpicture}[tdplot_main_coords, scale=0.7, baseline=2.5mm]
	% ボンド	
	\foreach \x in {1, 2, ..., 5} {
	\foreach \y in {1, 2} {
	\foreach \z in {0, 1} {
		\draw (\x, \y, 0) -- (\x, \y, 1) ;
		\draw (0.75, \y, \z) -- (5.25, \y, \z) ;
		\draw (\x, 0.75, \z) -- (\x, 2.25, \z) ;
	} } }
	% 生成演算子
	\node[fill, circle] at (2, 2, 1) {} ;
	% 消滅演算子
	\node[fill=white, draw, circle, inner sep=1mm, line width=0.5mm] at (3, 2, 1) {} ;
	% 個数演算子
	\node[fill=white, draw, inner sep=1.2mm, line width=0.5mm] at (3, 2, 0) {} ;
\end{tikzpicture}

\end{minipage}
\quad
\begin{minipage}{0.3\linewidth}
\center
$ \sfE _ { 3 } = $
\begin{tikzpicture}[tdplot_main_coords, scale=0.7, baseline=2.5mm]
	% ボンド	
	\foreach \x in {1, 2, ..., 5} {
	\foreach \y in {1, 2} {
	\foreach \z in {0, 1} {
		\draw (\x, \y, 0) -- (\x, \y, 1) ;
		\draw (0.75, \y, \z) -- (5.25, \y, \z) ;
		\draw (\x, 0.75, \z) -- (\x, 2.25, \z) ;
	} } }
	% 生成演算子
	\node[fill, circle] at (3, 2, 1) {} ;
	% 消滅演算子
	\node[fill=white, draw, circle, inner sep=1mm, line width=0.5mm] at (4, 2, 1) {} ;
	% 個数演算子
	\node[fill=white, draw, inner sep=1.2mm, line width=0.5mm] at (3, 2, 0) {} ;
\end{tikzpicture}

\end{minipage}\\
\vspace{1mm}
\begin{minipage}{0.3\linewidth}
\center
$ \sfF _ { 2 } = $
\begin{tikzpicture}[tdplot_main_coords, scale=0.7, baseline=2.5mm]
	% ボンド	
	\foreach \x in {1, 2, ..., 5} {
	\foreach \y in {1, 2} {
	\foreach \z in {0, 1} {
		\draw (\x, \y, 0) -- (\x, \y, 1) ;
		\draw (0.75, \y, \z) -- (5.25, \y, \z) ;
		\draw (\x, 0.75, \z) -- (\x, 2.25, \z) ;
	} } }
	% 生成演算子
	\node[fill, circle] at (2, 2, 1) {} ;
	\node[fill, circle] at (3, 2, 0) {} ;
	% 消滅演算子
	\node[fill=white, draw, circle, inner sep=1mm, line width=0.5mm] at (3, 2, 1) {} ;
	\node[fill=white, draw, circle, inner sep=1mm, line width=0.5mm] at (3, 1, 0) {} ;
\end{tikzpicture}

\end{minipage}
\quad
\begin{minipage}{0.3\linewidth}
\center
$ \sfF _ { 3 } = $
\begin{tikzpicture}[tdplot_main_coords, scale=0.7, baseline=2.5mm]
	% ボンド	
	\foreach \x in {1, 2, ..., 5} {
	\foreach \y in {1, 2} {
	\foreach \z in {0, 1} {
		\draw (\x, \y, 0) -- (\x, \y, 1) ;
		\draw (0.75, \y, \z) -- (5.25, \y, \z) ;
		\draw (\x, 0.75, \z) -- (\x, 2.25, \z) ;
	} } }
	% 生成演算子
	\node[fill, circle] at (3, 2, 1) {} ;
	\node[fill, circle] at (3, 2, 0) {} ;
	% 消滅演算子
	\node[fill=white, draw, circle, inner sep=1mm, line width=0.5mm] at (4, 2, 1) {} ;
	\node[fill=white, draw, circle, inner sep=1mm, line width=0.5mm] at (3, 1, 0) {} ;
\end{tikzpicture}

\end{minipage}\\
\vspace{1mm}
\begin{minipage}{0.3\linewidth}
\ 
\end{minipage}
\quad
\begin{minipage}{0.3\linewidth}
\center
$ \sfG _ { 3 } = $
\begin{tikzpicture}[tdplot_main_coords, scale=0.7, baseline=2.5mm]
	% ボンド	
	\foreach \x in {1, 2, ..., 5} {
	\foreach \y in {1, 2} {
	\foreach \z in {0, 1} {
		\draw (\x, \y, 0) -- (\x, \y, 1) ;
		\draw (0.75, \y, \z) -- (5.25, \y, \z) ;
		\draw (\x, 0.75, \z) -- (\x, 2.25, \z) ;
	} } }
	% 生成演算子
	\node[fill, circle] at (3, 2, 0) {} ;
	% 消滅演算子
	\node[fill=white, draw, circle, inner sep=1mm, line width=0.5mm] at (3, 1, 0) {} ;
	% 個数演算子
	\node[fill=white, draw, inner sep=1.2mm, line width=0.5mm] at (3, 2, 1) {} ;
\end{tikzpicture}

\end{minipage}
\vspace{3mm}
\captionsetup{width=0.8\linewidth}
\caption[dummy]{
The products $ \sfC _ { 1 } $, $ \sfE _ { j } $, $ \sfF _ { j } $ and $ \sfG _ { 3 } $ for $ \kb = 3 $.
Here, $ \sfF _ { 2 } $ is generated only by $ \sfE _ { 2 } $ and $ \sfG _ { 3 } $, $ \sfF _ { 3 } $ is generated only by $ \sfE _ { 3 } $ and $ \sfG _ { 3 } $.
}
\label{high dimensional effect case k = 3}
\end{figure}

We first note that because of Lemma 3.7, there are no products with $ \Wid = 3 = \kb $ with nonzero coefficients that generate $ \sfF _ { 2 } $.
There are several products with $ \Wid = 2 = \kb - 1 $ that generate $ \sfF _ { 2 } $, but Lemma 3.8 guarantees that $ \sfE _ { 2 } $ is the only one with possibly nonzero coefficient.
Finally $ \sfG _ { 3 } $ is the unique product with $ \Wid = 1 $ that generates $ \sfF _ { 2 } $.
We thus see that the coefficient \eqref{coefficient} for $ \sfF _ { 2 } $ is
\eq
r _ { \sfF _ { 2 } } = - t q _ { \sfE _ { 2 } } + t q _ { \sfG _ { 3 } }
\en
Similarly, we see that $ \sfE _ { 3 } $ and $ \sfG _ { 3 } $ are the only relevant products that generate $ \sfF _ { 3 } $, and hence
\eq
r _ { \sfF _ { 3 } } = - t q _ { \sfE _ { 3 } } - t q _ { \sfG _ { 3 } }
\en
Requiring $ r _ { \sfF _ { 2 } } = r _ { \sfF _ { 3 } } = 0 $ and recalling $ q _ { \sfE _ { 2 } } = q _ { \sfE _ { 3 } } $, we find $ q _ { \sfE _ { 2 } }  = 0 $, which is our goal.

\subsubsection{The case with $ \kb = 4 $}

As above, we shall define
\eqa
\sfE _ { j } &= \cp _ { ( j , 1 ) , \up } \n _ { ( 4 , 1 ) , \down } \cm _ { ( j + 2 , 1 ) , \up } , \\
\sfF _ { j } &= \cp _ { ( j , 1 ) , \up } \cp _ { ( 4 , 1 ) , \down } \cm _ { ( 4 , 2 ) , \down } \cm _ { ( j + 2 , 1 ) , \up } ,
\ena
for $ j = 2 , 3 $, and $ 4 $, and
\eqa
\sfG _ { 3 } &= \cp _ { ( 3 , 1 ) , \up } \cm _ { ( 4 , 1 ) , \up } \cp _ { ( 4 , 1 ) , \down } \cm _ { ( 4 , 2 ) , \down } , \qquad & \sfG _ { 4 } &= \cp _ { ( 4 , 1 ) , \up } \cp _ { ( 4 , 1 ) , \down } \cm _ { ( 4 , 2 ) , \down } \cm _ { ( 5 , 1 ) , \up } .
\ena
See Figure \ref{high dimensional effect case k = 4}.
\begin{figure}[htb]
\center
\begin{minipage}{0.3\linewidth}
\center
$ \sfC _ { 1 } = $
\begin{tikzpicture}[tdplot_main_coords, scale=0.7]

	% ボンド	
	\foreach \x in {1, 2, ..., 7} {
	\foreach \y in {1, 2} {
	\foreach \z in {0, 1} {
		\draw (\x, \y, 0) -- (\x, \y, 1) ;
		\draw (0.75, \y, \z) -- (7.25, \y, \z) ;
		\draw (\x, 0.75, \z) -- (\x, 2.25, \z) ;
	} } }
	
	% 生成演算子
	\node[fill, circle] at (1, 2, 1) {} ;
	% 消滅演算子
	\node[fill=white, draw, circle, inner sep=1mm, line width=0.5mm] at (4, 2, 1) {} ;
	
\end{tikzpicture}

\end{minipage}
\hfill
\begin{minipage}{0.3\linewidth}
\ 
\end{minipage}
\hfill
\begin{minipage}{0.3\linewidth}
\ 
\end{minipage}
\vspace{1mm}
\begin{minipage}{0.3\linewidth}
\center
$ \sfE _ { 2 } = $
\begin{tikzpicture}[tdplot_main_coords, scale=0.7]

	% ボンド	
	\foreach \x in {1, 2, ..., 7} {
	\foreach \y in {1, 2} {
	\foreach \z in {0, 1} {
		\draw (\x, \y, 0) -- (\x, \y, 1) ;
		\draw (0.75, \y, \z) -- (7.25, \y, \z) ;
		\draw (\x, 0.75, \z) -- (\x, 2.25, \z) ;
	} } }
	
	% 生成演算子
	\node[fill, circle] at (2, 2, 1) {} ;
	% 消滅演算子
	\node[fill=white, draw, circle, inner sep=1mm, line width=0.5mm] at (4, 2, 1) {} ;
	% 個数演算子
	\node[fill=white, draw, inner sep=1.2mm, line width=0.5mm] at (4, 2, 0) {} ;
	
\end{tikzpicture}

\end{minipage}
\hfill
\begin{minipage}{0.3\linewidth}
\center
$ \sfE _ { 3 } = $
\begin{tikzpicture}[tdplot_main_coords, scale=0.7]

	% ボンド	
	\foreach \x in {1, 2, ..., 7} {
	\foreach \y in {1, 2} {
	\foreach \z in {0, 1} {
		\draw (\x, \y, 0) -- (\x, \y, 1) ;
		\draw (0.75, \y, \z) -- (7.25, \y, \z) ;
		\draw (\x, 0.75, \z) -- (\x, 2.25, \z) ;
	} } }
	
	% 生成演算子
	\node[fill, circle] at (3, 2, 1) {} ;
	% 消滅演算子
	\node[fill=white, draw, circle, inner sep=1mm, line width=0.5mm] at (5, 2, 1) {} ;
	% 個数演算子
	\node[fill=white, draw, inner sep=1.2mm, line width=0.5mm] at (4, 2, 0) {} ;
	
\end{tikzpicture}

\end{minipage}
\hfill
\begin{minipage}{0.3\linewidth}
\center
$ \sfE _ { 4 } = $
\begin{tikzpicture}[tdplot_main_coords, scale=0.7]

	% ボンド	
	\foreach \x in {1, 2, ..., 7} {
	\foreach \y in {1, 2} {
	\foreach \z in {0, 1} {
		\draw (\x, \y, 0) -- (\x, \y, 1) ;
		\draw (0.75, \y, \z) -- (7.25, \y, \z) ;
		\draw (\x, 0.75, \z) -- (\x, 2.25, \z) ;
	} } }
	
	% 生成演算子
	\node[fill, circle] at (4, 2, 1) {} ;
	% 消滅演算子
	\node[fill=white, draw, circle, inner sep=1mm, line width=0.5mm] at (6, 2, 1) {} ;
	% 個数演算子
	\node[fill=white, draw, inner sep=1.2mm, line width=0.5mm] at (4, 2, 0) {} ;
	
\end{tikzpicture}

\end{minipage}
\vspace{1mm}
\begin{minipage}{0.3\linewidth}
\center
$ \sfF _ { 2 } = $
\begin{tikzpicture}[tdplot_main_coords, scale=0.7]

	% ボンド	
	\foreach \x in {1, 2, ..., 7} {
	\foreach \y in {1, 2} {
	\foreach \z in {0, 1} {
		\draw (\x, \y, 0) -- (\x, \y, 1) ;
		\draw (0.75, \y, \z) -- (7.25, \y, \z) ;
		\draw (\x, 0.75, \z) -- (\x, 2.25, \z) ;
	} } }
	
	% 生成演算子
	\node[fill, circle] at (2, 2, 1) {} ;
	\node[fill, circle] at (4, 2, 0) {} ;
	% 消滅演算子
	\node[fill=white, draw, circle, inner sep=1mm, line width=0.5mm] at (4, 2, 1) {} ;
	\node[fill=white, draw, circle, inner sep=1mm, line width=0.5mm] at (4, 1, 0) {} ;
	
\end{tikzpicture}

\end{minipage}
\hfill
\begin{minipage}{0.3\linewidth}
\center
$ \sfF _ { 3 } = $
\begin{tikzpicture}[tdplot_main_coords, scale=0.7]

	% ボンド	
	\foreach \x in {1, 2, ..., 7} {
	\foreach \y in {1, 2} {
	\foreach \z in {0, 1} {
		\draw (\x, \y, 0) -- (\x, \y, 1) ;
		\draw (0.75, \y, \z) -- (7.25, \y, \z) ;
		\draw (\x, 0.75, \z) -- (\x, 2.25, \z) ;
	} } }
	
	% 生成演算子
	\node[fill, circle] at (3, 2, 1) {} ;
	\node[fill, circle] at (4, 2, 0) {} ;
	% 消滅演算子
	\node[fill=white, draw, circle, inner sep=1mm, line width=0.5mm] at (5, 2, 1) {} ;
	\node[fill=white, draw, circle, inner sep=1mm, line width=0.5mm] at (4, 1, 0) {} ;
	
\end{tikzpicture}

\end{minipage}
\hfill
\begin{minipage}{0.3\linewidth}
\center
$ \sfF _ { 4 } = $
\begin{tikzpicture}[tdplot_main_coords, scale=0.7]

	% ボンド	
	\foreach \x in {1, 2, ..., 7} {
	\foreach \y in {1, 2} {
	\foreach \z in {0, 1} {
		\draw (\x, \y, 0) -- (\x, \y, 1) ;
		\draw (0.75, \y, \z) -- (7.25, \y, \z) ;
		\draw (\x, 0.75, \z) -- (\x, 2.25, \z) ;
	} } }
	
	% 生成演算子
	\node[fill, circle] at (4, 2, 1) {} ;
	\node[fill, circle] at (4, 2, 0) {} ;
	% 消滅演算子
	\node[fill=white, draw, circle, inner sep=1mm, line width=0.5mm] at (6, 2, 1) {} ;
	\node[fill=white, draw, circle, inner sep=1mm, line width=0.5mm] at (4, 1, 0) {} ;
	
\end{tikzpicture}

\end{minipage}\\
\vspace{1mm}
\begin{minipage}{0.3\linewidth}
\ 
\end{minipage}
\hfill
\begin{minipage}{0.3\linewidth}
\center
$ \sfG _ { 3 } = $
\begin{tikzpicture}[tdplot_main_coords, scale=0.7]

	% ボンド	
	\foreach \x in {1, 2, ..., 7} {
	\foreach \y in {1, 2} {
	\foreach \z in {0, 1} {
		\draw (\x, \y, 0) -- (\x, \y, 1) ;
		\draw (0.75, \y, \z) -- (7.25, \y, \z) ;
		\draw (\x, 0.75, \z) -- (\x, 2.25, \z) ;
	} } }
	
	% 生成演算子
	\node[fill, circle] at (3, 2, 1) {} ;
	\node[fill, circle] at (4, 2, 0) {} ;
	% 消滅演算子
	\node[fill=white, draw, circle, inner sep=1mm, line width=0.5mm] at (4, 2, 1) {} ;
	\node[fill=white, draw, circle, inner sep=1mm, line width=0.5mm] at (4, 1, 0) {} ;
	
\end{tikzpicture}

\end{minipage}
\hfill
\begin{minipage}{0.3\linewidth}
\center
$ \sfG _ { 4 } = $
\begin{tikzpicture}[tdplot_main_coords, scale=0.7]

	% ボンド	
	\foreach \x in {1, 2, ..., 7} {
	\foreach \y in {1, 2} {
	\foreach \z in {0, 1} {
		\draw (\x, \y, 0) -- (\x, \y, 1) ;
		\draw (0.75, \y, \z) -- (7.25, \y, \z) ;
		\draw (\x, 0.75, \z) -- (\x, 2.25, \z) ;
	} } }
	
	% 生成演算子
	\node[fill, circle] at (4, 2, 1) {} ;
	\node[fill, circle] at (4, 2, 0) {} ;
	% 消滅演算子
	\node[fill=white, draw, circle, inner sep=1mm, line width=0.5mm] at (5, 2, 1) {} ;
	\node[fill=white, draw, circle, inner sep=1mm, line width=0.5mm] at (4, 1, 0) {} ;
	
\end{tikzpicture}

\end{minipage}
\vspace{3mm}
\captionsetup{width=0.8\linewidth}
\caption[dummy]{
The products $ \sfC _ { 1 } $, $ \sfE _ { j } $, $ \sfF _ { j } $ and $ \sfG _ { j } $ for $ \kb = 4 $.
Here, $ \sfF _ { 2 } $ is generated only by $ \sfE _ { 2 } $ and $ \sfG _ { 3 } $, $ \sfF _ { 3 } $ is generated only by $ \sfE _ { 3 } $, $ \sfG _ { 3 } $ and $ \sfG _ { 4 } $, $ \sfF _ { 4 } $ is generated only by $ \sfE _ { 4 } $ and $ \sfG _ { 4 } $.
}
\label{high dimensional effect case k = 4}
\end{figure}
Exactly as in the case with $ \kb = 3 $, we obtain
\begin{align}
	r _ { \sfF _ { 2 } } &= - t q _ { \sfE _ { 2 } } + t q _ { \sfG _ { 3 } } , \\
	r _ { \sfF _ { 4 } } &= - t q _ { \sfE _ { 4 } } - t q _ { \sfG _ { 4 } } .
\end{align}
As for $ \sfF _ { 3 } $, we note that Lemma 3.8 implies $ \sfE _ { 3 } $ is the only product with $ \Wid = 3 $ and possibly nonzero coefficients that generates $ \sfF _ { 3 } $.
Clearly $ \sfG _ { 3 } $ and $ \sfG _ { 4 } $ are the only products with width 2 that generate $ \sfF _ { 3 } $.
We then find
\begin{align}
	r _ { \sfF _ { 3 } } &= - t q _ { \sfE _ { 3 } } - t q _ { \sfG _ { 3 } } + t q _ { \sfG _ { 4 } } .
\end{align}

Requiring that $ r _ { \sfF _ { j } } = 0 $, we get the set of equations
\begin{gather}
	- t q _ { \sfE _ { 2 } } + t q _ { \sfG _ { 3 } } = 0 , \\
	- t q _ { \sfE _ { 3 } } - t q _ { \sfG _ { 3 } } + t q _ { \sfG _ { 4 } } = 0 , \\
	- t q _ { \sfE _ { 4 } } - t q _ { \sfG _ { 4 } } = 0 ,
\end{gather}
which, with the constancy of $ q _ { \sfE _ { j } } $, implies $ q _ { \sfE _ { j } } = 0 $.

\subsubsection{The case with general $ \kb $}

The case with $ \kb $ such that $ 4 \leq \kb \leq \frac{ L }{ 2 } $ can be treated in essentially the same manner as the case with $ \kb = 4 $.

We define
\eqa
\sfE _ { j } &= \cp _ { ( j , 1 ) , \up } \n _ { ( \kb , 1 ) , \down } \cm _ { ( j + \kb - 2 , 1 ) , \up } , \\
\sfF _ { j } &= \cp _ { ( j , 1 ) , \up } \cp _ { ( \kb , 1 ) , \down } \cm _ { ( \kb , 2 ) , \down } \cm _ { ( j + \kb - 2 , 1 ) , \up } ,
\ena
for $ j = 2 , \ldots , \kb $, and
\eqa
\sfG _ { j } &= \cp _ { ( j , 1 ) , \up } \cp _ { ( \kb , 1 ) , \down } \cm _ { ( \kb , 2 ) , \down } \cm _ { ( j + \kb - 3 , 1 ) , \up } ,
\ena
for $ j = 3 , \ldots , \kb $, where $ \Wid \sfE _ { j } = \Wid \sfF _ { j } = \kb - 1 $ and $ \Wid \sfG _ { j } = \kb - 2 $.

One then finds that the coefficients for $ \sfF _ { j } $ are given by
\begin{align}
	r _ { \sfF _ { 2 } } &= - t q _ { \sfE _ { 2 } } + t q _ { \sfG _ { 3 } } , \label{coefficient in proof as general k 1} \\
	r _ { \sfF _ { \kb } } &= - t q _ { \sfE _ { \kb } } - t q _ { \sfG _ { \kb } } , \label{coefficient in proof as general k 2}
\end{align}
and for $ j = 3 , \ldots , \kb - 1 $
\begin{align}
	r _ { \sfF _ { j } } &= - t q _ { \sfE _ { j } } - t q _ { \sfG _ { j } } + t q _ { \sfG _ { j + 1 } } . \label{coefficient in proof as general k 3}
\end{align}
By demanding $ r _ { \sfD _ { j } } = 0 $ and recalling that $ q _ { \sfE _ { j } } $ is independent of $ j $, we get $ q _ { \sfE _ { j } } = 0 $.

\section{Discussion}
\label{s:discussion}
We studied the standard Hubbard model with Hamiltonian \eqref{Hamiltonian} defined on the $d$-dimensional hypercubic lattice with $d\ge2$.
We proved that the model admits no nontrivial local conserved quantities provided that $ U \neq 0 $ and $ t \neq 0 $. 
The absence of nontrivial local conserved quantities strongly suggests that the model is non-integrable, in contrast to its one-dimensional counterpart.

\medskip
As we have stressed in Section~\ref{s:intro} and at the end of Section~\ref{s:strategy}, our proof is {\em not}\/ a straightforward exntension of that by Shiraishi and Tasaki \cite{ShiraishiTasaki2024}, who proved a similar theorem for the $S=\tfrac12$ XY and XYZ spin models in $d\ge2$.
The proof for the Hubbard model is more delicate and requires an extra step.
Roughly speaking, the difficulty in the Hubbard model comes from the fact that the free fermion model obtained by setting $U=0$ in \eqref{Hamiltonian} is integrable in any dimension, and the fact that the one-dimensional Hubbard model is integrable.
A legitimate proof must take into account both the high-dimensionality and the nonzero $U$.

Lemma~\ref{creation-annihilation pair} showed that the products with the maximum width $\kb$ in a candidate of conserved quantity have the standard form \eqref{standard form0}.
This simple form, consisting of an annihilation and a creation operator, may be regarded as a manifestation of the integrability of the free fermion model.
We note that in the corresponding proof, say in \cite{ShiraishiTasaki2024}, for quantum spin models, a close analysis of the products with the maximum width is essentially sufficient to complete the proof of the absence of nontrivial local conserved quantities.
In the Hubbard model, on the other hand, we get little information from the products \eqref{standard form0} with the maximum width.
This is why we have to go ``one step further'' and prove Lemma~\ref{substandard form} to restrict the form of products with the next maximum width.

As we discussed below Lemma~\ref{substandard form}, it was necessary for our proof to partially specify the local conserved quantities for the integrable one-dimensional Hubbard model.
This is in stark contrast with the proof in \cite{ShiraishiTasaki2024}; it equally applies to the $S=\tfrac12$ XY and XYZ models with or without a magnetic field, independent of the exact form (or even the presence/absence) of conserved quantities in the corresponding one-dimensional models.
See also footnote~\ref{fn:Shiraishi}.

\medskip
In the present paper, we only treated the standard Hubbard Hamiltonian \eqref{Hamiltonian} with an isotropic hopping amplitude.
Our proof automatically extends to models with nearest neighbor hopping whose amplitude depends on the direction.
Although we treated real hopping amplitude, mainly for notational simplicity, it is also possible to treat complex hopping amplitude.
Interestingly our proof covers Hamiltonians that are not even Hermitian.
Finally, as is clear from our diagramatic representations, our proof does not require a full $d$-dimensional hypercubic lattice.
As in \cite{ShiraishiTasaki2024,Chiba2024}, the proof of the absence of conserved quantities works for the Hubbard model defined on a ladder.

We also expect that our method can be extended to prove the absence of nontrivial local conserved quantities in other lattice fermion models of physical interest.

\appendix

\section{Appendix: Models with hopping amplitudes that are not real symmetric}

Let us consider the model in which the hopping Hamiltonian \eqref{hopping term} is generalized as
\eq
\Hhop ^ { \p } = - \sum _ { x \in \L } \sum _ { \a = 1 , \ldots , d } \sum _ { \s = \up , \down } \Big ( t _ { - } \cp _ { x , \s } \cm _ { x + e _ { \a } , \s } + t _ { + } \cp _ { x + e _ { \a } , \s } \cm _ { x , \s } \Big ) , \label{hopping term in appendix}
\en
where $ t _ { \pm } \in \mathbb{C} \setminus \{ 0 \} $.
We do not assume that $ t _ { - } $ and $ t _ { + } $ are complex conjugates of each other.
This means that we treat hopping Hamiltonians that are not necessarily Hermitian.
When $ t _ { - } = t _ { + } = t \in \mathbb{ R } $, equation \eqref{hopping term in appendix} coincides with \eqref{hopping term}.

With the hopping Hamiltonian \eqref{hopping term in appendix}, equation \eqref{E:coefficient of Shift} in Lemma \ref{L:two site product} is modified as
\eq
\qSA = - \a \b \frac{ t _ { \b } }{ t _ { \bo } } \qA .
\en
Moreover, in Lemmas \ref{L:horizontal condition}--\ref{L:same spin}, the proofs are altered in the sense that equations \eqref{E:first coefficient in the proof of Lemma 3.5}--\eqref{E:third coefficient in the proof of Lemma 3.5} are replaced by the following, though the content of the lemmas remains unchanged:
\eqa
r _ { \sfD _ { 1 } ^ { \p } } &= - \b U q _ { \sfA } + \a t _ { \ao } q _ { \sfE _ { 2 } ^ { \p } } , \\
r _ { \sfD _ { l } ^ { \p } } &= \b t _ { \b } q _ { \sfE _ { l } ^ { \p } } + \a t _ { \ao } q _ { \sfE _ { l + 1 } ^ { \p } } , \quad l = 2 , \ldots , \kb - 1 , \\
r _ { \sfD _ { \kb } ^ { \p } } &= \b t _ { \b } q _ { \sfE _ { \kb } ^ { \p } } .
\ena
Again, by noting $ r _ { \sfD _ { j } ^ { \p } } = 0 $, we similarly find $ \qA = 0 $.

The modifications required in the proof of Lemma \ref{creation-annihilation pair} demand somewhat more care. First, equation \eqref{coefficient from Lemma 3.4} is modified as follows:
\eq
q _ { \sfC _ { \kb } ^ { \p \p \p } } = ( - 1 ) ^ { \kb - 1 } \Big ( \frac{ t _ { + } }{ t _ { - } } \Big ) ^ { \kb - 1 } q _ { \sfC _ { 1 } ^ { \p \p \p } } . \label{coefficient of k-times shift1}
\en
Furthermore, equations \eqref{E:first coefficient in the proof of Lemma 3.7}--\eqref{E:third coefficient in the proof of Lemma 3.7} are replaced by the following:
\eqa
r _ { \sfD _ { 1 } ^ { \p \p \p } } &= - U q _ { \sfC _ { 1 } ^ { \p \p \p } } + t _ { - } q _ { \sfE _ { 2 } ^ { \p \p \p } } , \\
r _ { \sfD _ { l } ^ { \p \p \p } } &= t _ { + } q _ { \sfE _ { l } ^ { \p \p \p } } + t _ { - } q _ { \sfE _ { l + 1 } ^ { \p \p \p } } , \qquad l = 2 , \ldots , \kb - 1 , \\
r _ { \sfD _ { \kb } ^ { \p \p \p } } &= t _ { + } q _ { \sfE _ { \kb } ^ { \p \p \p } } - U q _ { \sfC _ { \kb } ^ { \p \p \p } } .
\ena
Again noting $ r _ { \sfD _ { j } ^ { \p } } = 0 $, we obtain
\eq
q _ { \sfC _ { \kb } ^ { \p \p \p } } = ( - 1 ) ^ { \kb - 2 } \Big ( \frac{ t _ { + } }{ t _ { - } } \Big ) ^ { \kb - 1 } q _ { \sfC _ { 1 } ^ { \p \p \p } } . \label{coefficient of k-times shift2}
\en
Comparing \eqref{coefficient of k-times shift1} and \eqref{coefficient of k-times shift2}, we find $ q _ { \sfC _ { 1 } ^ { \p \p \p } } = 0 $.

In addition, equations \eqref{coefficient in proof as general k 1}--\eqref{coefficient in proof as general k 3} in the Third step are replaced by the following:
\eqa
r _ { \sfF _ { 2 } } &= - t _ { - } q _ { \sfE _ { 2 } } + t _ { - } q _ { \sfG _ { 3 } } , \\
r _ { \sfF _ { j } } &= - t _ { - } q _ { \sfE _ { j } } - t _ { - } q _ { \sfG _ { j } } + t _ { - } q _ { \sfG _ { j + 1 } } , \qquad j = 3 , \ldots , \kb - 1 , \\
r _ { \sfF _ { \kb } } &= - t _ { - } q _ { \sfE _ { \kb } } - t _ { - } q _ { \sfG _ { \kb } } .
\ena
As before, these lead to the conclusion that $ q _ { \sfE _ { j } } = 0 $.

Although these modifications arise in the proofs of Steps 1–3 as a result of the altered hopping term, the overall conclusions remain unaffected.
From this, we conclude that even with the generalized (possibly non-Hermitian) hopping Hamiltonian \eqref{hopping term in appendix}, the higher-dimensional Hubbard model admits no nontrivial local conserved quantities.

\bigskip
\noindent{\small
{\em Acknowledgement:} \/
I would like to thank Hal Tasaki for suggesting the problem, his invaluable discussions, and his careful reading of the manuscript, Akihiro Hokkyo and Mizuki Yamaguchi for pointing out errors in the earlier version of the manuscript and for the stimulating discussions, and Kohei Fukai and Kanji Yamada for their valuable comments.
I also thank Naoto Shiraishi for providing helpful information regarding the proof of the absence of local conserved quantities in various models, and Hosho Katsura for pointing me to some useful references.
The present work is supported in part by JSPS Grants-in-Aid for Scientific Research No.~25K07171.
}


\begin{thebibliography}{99}

\bibitem{Lieb1995}
E. H. Lieb,
{\em The Hubbard Model: Some Rigorous Results and Open Problems}\/,
XI Int. Cong. MP, Int. Press, 392--412 (1995).
\\\url{https://arxiv.org/abs/cond-mat/9311033}

\bibitem{Tasaki1995}
H. Tasaki,
{\it  The Hubbard Model — Introduction and Selected Rigorous Results}\/,
J. Phys.: Condens. Matter \textbf{10} 4353 (1998).
\\\url{https://arxiv.org/abs/cond-mat/9512169}

\bibitem{Arovas}
D. P. Arovas, E. Berg, S. Kivelson, and S. Raghu,
{\em The Hubbard Model}\/,
Annual Reviews of Condensed Matter Physics {\bf 13}, 239 (2022).
\\\url{https://arxiv.org/abs/2103.12097}

\bibitem{LiebWu1968}
E. H. Lieb and F. Y. Wu,
{\it Absence of Mott Transition in an Exact Solution of the Short-Range, One-Band Model in One Dimension}\/,
Phys. Rev. Lett. \textbf{20}, 1445 (1968).
\\\url{https://journals.aps.org/prl/abstract/10.1103/PhysRevLett.20.1445}

\bibitem{LiebWu2003}
E. H. Lieb and F. Y. Wu,
{\it The one-dimensional Hubbard model: A reminiscence}\/,
Physica A {\bf 321}, 1--27 (2003).
\\\url{https://arxiv.org/abs/cond-mat/0207529}

\bibitem{Shastry1986}
B. S. Shastry,
{\it Infinite Conservation Laws in the One-Dimensional Hubbard Model}\/,
Phys. Rev. Lett. \textbf{56}, 1529 (1986).
\\\url{https://journals.aps.org/prl/abstract/10.1103/PhysRevLett.56.1529}

\bibitem{Shastry19867}
B. S. Shastry,
{\it Exact Integrability of the One-Dimensional Hubbard Model}\/,
Phys. Rev. Lett. \textbf{56}, 2453 (1986).
\\\url{https://journals.aps.org/prl/abstract/10.1103/PhysRevLett.56.2453}

\bibitem{Shastry1988}
B. S. Shastry,
{\it Decorated Star-Triangle Relations and Exact Integrability of the One-Dimensional Hubbard Model}\/,
J. Stat. Phys. \textbf{50}(1), 57--79 (1988). 
\\\url{https://link.springer.com/article/10.1007/BF01022987}
	
\bibitem{Grosse1989}
H. Grosse,
{\it The symmetry of the Hubbard model}\/,
Lett Math Phys \textbf{18}, 151--156 (1989).
\\\url{https://doi.org/10.1007/BF00401869}
	
\bibitem{Yang1989}
C. N. Yang,
{\it $ \eta $ Pairing and Off-Diagonal Long-Range Order in a Hubbard Model}\/,
Phys. Rev. Lett. \textbf{63}, 2144 (1989).
\\\url{https://doi.org/10.1103/PhysRevLett.63.2144}
	
\bibitem{GrabowskiMathieu1994}
M. P. Grabowski, and P. Mathieu,
{\it Structure of the conservation laws in integrable spin chains with short range interactions}\/,
Annals Phys. {\bf 243},  299--371(1995).
\\\url{https://arxiv.org/abs/hep-th/9411045v1}
	
\bibitem{Fukai2023}
K. Fukai,
{\it All local conserved quantities of the one-dimensional Hubbard model}\/,
Phys. Rev. Lett. \textbf{131}, 256704 (2023).
\\\url{https://arxiv.org/abs/2301.03621}
	
\bibitem{Fukai2024}
K. Fukai,
{\it Proof of completeness of the local conserved quantities in the one-dimensional Hubbard model}\/,
J. Stat. Phys. \textbf{191}, 70 (2024).
\\\url{https://arxiv.org/abs/2309.09354}

\bibitem{Shiraishi2019}
N. Shiraishi,
{\it Proof of the absence of local conserved quantities in the XYZ chain with a magnetic field}\/,
Europhys. Lett. \textbf{128} 17002 (2019).
\\\url{https://arxiv.org/abs/1803.02637}
	
\bibitem{Chiba2023}
Y. Chiba,
{\it Proof of absence of local conserved quantities in the mixed-field Ising chain}\/,
Phys. Rev. B \textbf{109}, 035123 (2023).
\\\url{https://arxiv.org/abs/2307.16703}
	
\bibitem{ParkLee2025}
H.K. Park and S. Lee,
{\it Graph-theoretical proof of nonintegrability in quantum many-body systems: Application to the PXP model}\/,
Phys. Rev. B \textbf{111}, L081101 (2025).
\\\url{https://arxiv.org/abs/2403.02335}
	
\bibitem{Shiraishi2024}
N. Shiraishi,
{\it Absence of Local Conserved Quantity in the Heisenberg Model with Next Nearest-Neighbor Interaction}\/,
J. Stat. Phys. \textbf{191}:114 (2024).
\\\url{https://link.springer.com/article/10.1007/s10955-024-03326-4}
	
\bibitem{ParkLee2024}
H.K. Park and S. Lee,
{\it Proof of nonintegrability of the spin-1 bilinear-biquadratic chain model}\/,
(preprint, 2024).
\\\url{https://arxiv.org/abs/2410.23286}
	
\bibitem{HokkyoYamaguchiChiba2024}
A. Hokkyo, M. Yamaguchi, and Y. Chiba,
{\it Proof of the absence of local conserved quantities in the spin-1 bilinear-biquadratic chain and its anisotropic extensions}\/,
(preprint, 2024).
\\\url{https://arxiv.org/abs/2411.04945}
	
\bibitem{YamaguchiChibaShiraishi2024a}
M. Yamaguchi, Y. Chiba, and N. Shiraishi,
{\it Complete Classification of Integrability and Non-integrability for Spin-1/2 Chain with Symmetric Nearest-Neighbor Interaction}\/,
(preprint, 2024).
\\\url{https://arxiv.org/abs/2411.02162}
	
\bibitem{YamaguchiChibaShiraishi2024b}
M. Yamaguchi, Y. Chiba, and N. Shiraishi,
{\it Proof of the absence of local conserved quantities in general spin-1/2 chains with symmetric nearest-neighbor interaction}\/,
(preprint, 2024).
\\\url{https://arxiv.org/abs/2411.02163}
	
\bibitem{Shiraishi2025}
N. Shiraishi,
{\it Complete classification of integrability and non-integrability of S=1/2 spin chains with symmetric next-nearest-neighbor interaction}\/,
(preprint, 2025).
\\\url{https://arxiv.org/abs/2501.15506}
	
\bibitem{Hokkyo2025}
A. Hokkyo,
{\it Rigorous Test for Quantum Integrability and Nonintegrability}\/,
(preprint, 2025).
\\\url{https://arxiv.org/abs/2501.18400}

\bibitem{ShiraishiYamaguchi2025}
N. Shiraishi and M. Yamaguchi,
{\em Dichotomy theorem distinguishing non-integrability and the lowest-order Yang-Baxter equation for isotropic spin chains}\/,
(preprint, 2025).
\\\url{https://arxiv.org/abs/2504.14315}
	
\bibitem{ShiraishiTasaki2024}
N. Shiraishi and H. Tasaki,
{\it The S = 1/2 XY and XYZ models on the two or higher dimentsional hypercubic lattice do not possess nontrivial local conserved quantities}\/,
(preprint, 2024).
\\\url{https://arxiv.org/abs/2412.18504}
	
\bibitem{Chiba2024}
Y. Chiba,
{\it Proof of absence of local conserved quantities in two- and higher-dimensional quantum Ising models}\/,
(preprint, 2024).
\\\url{https://arxiv.org/abs/2412.18903}
			
\bibitem{FutamiTasaki2025}
M. Futami and H. Tasaki,
{\it Absence of nontrivial local conserved quantities in the quantum compass model on the square lattice}\/,
(preprint, 2025).
\\\url{https://arxiv.org/abs/2502.10791v1}

\end{thebibliography}
\end{document}